\documentclass[paper, 12pt, letterpaper, epsf]{JHEP} \input{epsf.tex}
\usepackage{graphics} \usepackage{epsfig}
\usepackage{amsmath}
\usepackage{amssymb}

\def\Id{{\bf 1}}

\def\C{{\mathbb C}} \def\R{{\Bbb R}} \def\Z{{\Bbb Z}}
  \def\P{{\Bbb P}}

 \def\tr{\operatorname{tr}}

\def\tr{{\rm tr}} \def\W{{\cal W}}

\def\id{\protect{{1 \kern-.28em {\rm l}}}}

\newcommand{\be}{\begin{equation}} \newcommand{\ee}{\end{equation}}
\newcommand{\bea}{\begin{eqnarray}} \newcommand{\eea}{\end{eqnarray}}
\newcommand{\beann}{\begin{eqnarray*}}
  \newcommand{\eeann}{\end{eqnarray*}}
\newcommand{\bfig}{\begin{figure}} \newcommand{\efig}{\end{figure}}
\newcommand{\nn}{\nonumber}
\newcommand{\ba}{\begin{array}}\newcommand{\ea}{\end{array}}

\newtheorem{Proposition}{Proposition}[section]

\newtheorem{Theorem}{Theorem}[section]
\newtheorem{Lemma}{Lemma}[section]
\newtheorem{Corrolary}{Corrolary}[section]

\newcommand{\bp}{\begin{Proposition}}
  \newcommand{\ep}{\end{Proposition}}
\newcommand{\bt}{\begin{Theorem}} \newcommand{\et}{\end{Theorem}}
\newcommand{\bl}{\begin{Lemma}} \newcommand{\el}{\end{Lemma}}
\newcommand{\bc}{\begin{Corrolary}} \newcommand{\ec}{\end{Corrolary}}

\title{(Anti)symmetric matter and superpotentials from IIB
  orientifolds}

\author{K. Landsteiner\\Instituto de Fisica Theorica
  C-XVI\\Universidad Autonoma de Madrid\\28049 Madrid,
  Spain\\KarlLandsteiner@uam.es}
\author{C. I. Lazaroiu\\Humboldt
  Universit\"at zu Berlin\\ Newtonstrasse 15, 12489 Berlin-Adlershof, Germany\\calin@physik.hu-berlin.de}
\author{Radu Tatar\\
  Theoretical Physics Group \\
  Lawrence Berkeley National Laboratory\\
  Berkeley, CA 94720, USA\\rtatar@socrates.berkeley.edu
}

\abstract{We study the IIB engineering of {\cal N}=1
  gauge theories with unitary gauge group and matter in
  the adjoint and (anti)symmetric
  representations. We show that such theories can be obtained as $\Z_2$
  orientifolds of certain Calabi-Yau $A_2$ fibrations, and discuss 
  the explicit T-duality transformation to an orientifolded Hanany-Witten construction. 
  The low energy dynamics is
  described by a geometric transition of the orientifolded background.
  Unlike previously studied cases, we show that the orientifold
  5-`plane' survives the transition, thus bringing a nontrivial
  contribution to the effective superpotential.  We extract this
  contribution by using matrix model results and compare with 
  geometric data. A Higgs branch of our models
  recovers the engineering of SO/Sp 
  theories with adjoint matter through an O5-`plane' T-dual to
  an O6-plane. We show that the superpotential agrees with
  that produced by engineering through an O5-`plane' dual
  to an O4-plane, even though the orientifold of this second
  construction is replaced by fluxes after the transition.}

\preprint{HU-EP-03/30,
  IFT-UAM/CSIC-03-15\\
  LBNL-53008,~UCB-PTH-03/12 }
\begin{document}

\tableofcontents

\pagebreak

\vskip .6in

\section{Introduction}
\label{intro}

D-brane physics allows for a description of supersymmetric gauge
theories leading to novel insights into strong
coupling behavior.  A particularly fruitful approach in this regard
is afforded by geometric engineering.  This leads one to consider
D-branes partially wrapped on cycles of a nontrivial
geometry, which realizes the supersymmetric gauge theory of interest
on the uncompactified remnant of the branes' worldvolume.

A prominent example is given by certain local conifold geometries
which implement the ``large N geometric transitions''
\cite{vafa,civ, Cachazo_Vafa, Cachazo_Vafa_more}.  The starting point
of this construction is a IIB background whose closed string sector is
described by a singular ADE fibration $X_0$ over a complex plane
parameterized by $z$. The ADE fiber of $X_0$ degenerates above certain points
of the plane, where the total space acquires conifold singularities.
The small resolution ${\hat X}$ of $X_0$ contains a set of
holomorphically embedded two-spheres on which one can wrap D5-branes.
Wrapping $N_i$ D5-branes on the $i$-th exceptional $\P^1$ leads to a
four-dimensional ${\cal N}=1$ quiver gauge theory on the
uncompactified part of the branes' worldvolume. Such theories can be
viewed as softly broken ${\cal N}=2$ quiver gauge theories. The
partial supersymmetry breaking is induced by a superpotential for those
chiral multiplets which transform in the adjoint representation of the
gauge group. The precise form of the superpotential is determined by
the fibration data.

With a nontrivial superpotential, such theories confine at low
energies and lead to gaugino condensation. In the geometric
realization, this corresponds to a transition in which the exceptional
$\P^1$'s shrink to zero size and the resolved geometry is replaced by
a `log-normalizable' smoothing $X$ of the singular fibration $X_0$.
Thus each exceptional $\P^1$ is replaced by a 3-sphere. In this
process the D5-branes disappear but their RR-flux is still present and
supported on the three-cycles of the deformed geometry. This
represents a type IIB background with nontrivial 3-form fluxes and
therefore leads to a superpotential of the form \cite{tv}: \be
W_{eff} = \int H \wedge \Omega = \sum_i \left[N_i\frac{\partial
    F_0}{\partial S_i} + 2 \pi i \alpha_i S_i \right]\,,
\label{flux_superpotential} \ee where $\Omega$ is the holomorphic
three-form of the deformed Calabi-Yau $X$, $H$ is the type IIB three
form field strength, $S_i$ are the gaugino condensates, $N_i$ the
RR-fluxes and $\alpha_i$ the gauge couplings (which are identified
with the NS part of the $H$-flux). Here $F_0$ is the prepotential of
the closed string sector.  More precisely, one has: \bea
\label{identif}
2 \pi i S_i = \int_{A_i} \Omega&,& \frac{\partial F_0}{\partial S_i} = \Pi_i:=\int_{B_i} \Omega\\
N_i = \int_{A_i} H&,& \alpha_i = \int_{B_i} H~~, \eea
where $A_i, B_i$ give a symplectic basis of a nonstandard version of
$H^3(X,\Z)$ (note that the 
$B$ `cycles' are non-compact and must be regularized \footnote{
A way to work with {\em standard} compact cohomology 
is provided by the geometric regularization discussed in
\cite{matrix_reg}.}).  Many examples of this construction have
been studied in detail in recent years
\cite{Cachazo_Vafa,Cachazo_Vafa_more,OT}, including the addition of certain types of orientifolds
\cite{eot,ookouchi} as well as of fundamental matter which leads to mesonic
branches \cite{Radu_fundamentals1} and baryons \cite{Radu_fundamentals2}.
A T-dual  approach to such transitions was developed in \cite{OT1,OT2}, where relations
(\ref{identif}) appear naturally by studying the low energy dynamics of an M5
brane in MQCD. The MQCD approach \cite{wit1997} is based on lifting
a T-dual type IIA configuration with NS5 branes and D4 branes to M-theory.

The purpose of the present paper is to discuss a modification of the
geometric transitions of \cite{vafa,civ, Cachazo_Vafa,
  Cachazo_Vafa_more} which allows us to describe the low energy
dynamics of ${\cal N}=1$ $U(N)$ gauge theories with one adjoint chiral
multiplet and two additional chiral multiplets transforming in either the
symmetric or antisymmetric two-tensor representation and its conjugate. The tree-level
superpotential will have the form: \be \label{symasym_superpot}
W_{tree} = \tr\left[W(\Phi) + \tilde Q
\Phi Q \right] \ee where $\Phi$ is the adjoint chiral superfield
while the symmetric/antisymmetric chiral superfields $Q$ and ${\tilde Q}$ transform as $Q\rightarrow U Q U^T$ and
$\tilde Q\rightarrow U^* \tilde Q U^\dagger$.  These fields obey $Q^T=sQ$ and  ${\tilde Q}^T= s
{\tilde Q}$, where $s=\pm 1$. Here 
$W(\Phi) = \sum_{k=0}^d \frac{g_k}{k+1}
\tr ( \Phi^{k+1})$ is a polynomial of degree $d+1$. 
Such theories can be realized in the
type IIA set-up though orientifolded Hanany-Witten constructions
\cite{Karl_nonchiral0, Karl_nonchiral1} and have recently been
reconsidered in the context of the Dijkgraaf-Vafa conjecture \cite{us,
  Naculich, Naculich1}. The IIB approach discussed in this paper will
allow us to give the geometric engineering of such theories, which has
been missing until now.

As we shall see below, the IIB realization of such systems requires
an $A_2$ fibration together with a $\Z_2$ orientifold . 
Unlike the cases studied in \cite{ashok} (which
allow one to engineer the $SO(N)$ and $Sp(N/2)$ gauge theories with
one adjoint chiral multiplet), the orientifold action we consider
leads to an orientifold five-`plane' which {\em survives} the large
$N$ transition. This gives a string-theoretic explanation of the subtle
behavior of such theories which was extracted in \cite{us} and  
\cite{Naculich, Naculich1} in the context of the
matrix-model conjecture of Dijkgraaf and Vafa \cite{DV,DV2,DV3} as
well as through the method of generalized 
Konishi anomalies \cite{Witten1}.

The paper is organized as follows. In Section \ref{geom} we geometrically
engineer this class of gauge theories. The crucial ingredient is
the addition of an orientifold 5-`plane' to the resolution of a
nontrivial $A_2$ fibration.  We also discuss explicitly the T-duality
which maps our geometry to the orientifolded Hanany-Witten construction \cite{hw},
thus explaining the relation between our IIB construction and the IIA
description of \cite{Karl_nonchiral0, Karl_nonchiral1}.

Section \ref{transition} considers the geometric transition for such
models. 
Since our O5 `plane' survives the transition, it will contribute to the effective
superpotential.  We extract a geometric expression for this contribution by using the matrix
model results of \cite{us} and compare with the proposal of
\cite{Vafa_or}.

Section \ref{sosp} gives a comparative treatment of theories with orthogonal
or symplectic gauge group and adjoint matter, which can be engineered as orientifolds of
$A_1$ fibrations by using two different choices of O5 `planes'.
The first construction uses an O5-`plane' which is T-dual to an
O4-plane, and corresponds to the engineering considered in
\cite{ashok}. The internal part of this 5-`plane' is a compact 
exceptional curve (a $\P^1$) of the resolved fibration.  The second construction uses an O5-`plane'
whose internal part is a noncompact curve. This is T-dual to an
O6-plane in the Hanany-Witten construction. After the geometric
transition, the first construction leads to a pure flux background,
with the orientifold being replaced by a contribution to the R-R flux.
For the second construction, the O5-`plane' survives the transition
and thus contributes to the effective superpotential. 

Since both constructions 
engineer the same field theory, the effective superpotentials must
agree.  We check this agreement by showing that the 
second construction can be obtained by considering a certain Higgs
branch of the U(N) field theory with symmetric or antisymmetric
matter, which allows us to extract the flux-orientifold superpotential by using
the results of \cite{us}. This allows us to show that the spectral
curves of the associated matrix models agree between the two
constructions.  Our conclusions are summarized in Section \ref{conclusions}.

In Appendix \ref{without}, we turn off the superpotential
for the adjoint chiral multiplet. We describe the orientifold of the toric
resolution of the $A_2$ singularity and show how the resulting
O5 `plane' indeed gives rise to matter in the symmetric or antisymmetric
representation.

\section{Geometric engineering with a tree-level superpotential}
\label{geom}

We start by discussing the geometric engineering of our field theory
as an orientifold of a type IIB background with D5-branes. Without
the orientifold, this coincides with the background used to engineer
the ${\cal N}=1$ $A_2$ quiver field theory \cite{Cachazo_Vafa, Cachazo_Vafa_more}
and we start by recalling the latter.

\subsection{Geometric engineering of the $A_2$ quiver theory}

\subsubsection{The IIB background}

Let us consider IIB string theory on the resolution ${\hat X}$
of a non-compact Calabi-Yau threefold $X_0$ given by a singular $A_2$
fibration over the complex plane.  The background includes a
collection of D5-branes wrapping the exceptional $\P^1$'s of the
resolution.

Explicitly, the singular space $X_0$ can be realized as the
hypersurface: \be
\label{X0}
xy=(u-t_0(z))(u-t_1(z))(u-t_2(z))~~, \ee where $x,y,u,z$ are the
affine coordinates of $\C^4$ and the polynomials $t_j(z)$ are given
by: \bea\label{A2_fibration}
t_0(z)&:=&-\frac{2W'_1(z)+W'_2(z)}{3}\nn\\
t_1(z)&:=&\frac{2W'_2(z)+W'_1(z)}{3}\\
t_2(z)&:=&-t_0(z)-t_1(z)=\frac{W'_1(z)-W'_2(z)}{3}\nn~~.  \eea

Generically, the affine variety (\ref{X0}) has $A_1$ singularities at
$x=y=0$ and $z$ equal to one of the double points of the planar
algebraic curve: \be
\label{Sigma_0}
\Sigma_0:~~(u-t_0(z))(u-t_1(z))(u-t_2(z))=0~~.  \ee This curve gives a
reducible 3-section of the $A_2$ fibration (\ref{X0}) whose three
rational components $C_j$ are given by $u=t_j(z)$. The double points
of $\Sigma_0$ sit at the intersection of two such components. These
are obtained when $u=t_i(z)=t_j(z)$ for $i\neq j$, which gives the
equations: \bea
t_0(z)-t_2(z)&=&-W'_1(z)=0,~~u=-\frac{W'_2(z)}{3}\nn\\
t_1(z)-t_2(z)&=&+W'_2(z)=0,~~u=\frac{W'_1(z)}{3}\\
t_0(z)-t_1(z)&=&-W'_1(z)-W'_2(z)=0,~~u=-\frac{W'_1(z)}{3}=\frac{W'_2(z)}{3}\nn~~.
\eea We let $z_j^{(\alpha)}$ be the roots of $W'_\alpha(z)$, ${\tilde
  z}_j$ the roots of $W'_1(z)+W'_2(z)$, and denote the corresponding
exceptional $\P^1$'s of the resolution ${\hat X}$ by $D_j^{(\alpha)}$
and ${\tilde D}_j$ respectively. Throughout the rest of Section 2, we
assume that $W'_1(z)$ and $W'_2(z)$ have no common zeroes (which is
the generic situation).  This means that there is no point in the
$(z,u)$-plane where all three components $C_j$ intersect, i.e. the
curve $\Sigma_0$ does not have any {\em triple} points. With this
assumption, the sets $\{z_j^{(1)}\}$, $\{z^{(2)}_j\}$ and $\{{\tilde
  z}_j\}$ are mutually disjoint. In particular, we have
$W'_\alpha(z^{(\beta)}_j)\neq 0$ for $\alpha\neq \beta$ and
$W'_\alpha({\tilde z}_j)\neq 0$.  Thus the singularities of $X_0$ are
ordinary double points of the fibers $X_0(z_j^{(\alpha)})$ and
$X_0({\tilde z}_j)$. Then ${\hat X}$ is obtained by blowing up each of
these double points, thus replacing the singular fibers with their
minimal resolutions ${\hat X}(z_j^{(\alpha)})$ and ${\hat X}({\tilde
  z}_j)$.

The resolved space ${\hat X}$ can be described explicitly as
follows\footnote{Our coordinates differ from those of \cite{OT} by the
  shift $u\rightarrow u-t_0(z)$.}  \cite{OT}. Consider two copies of
$\P^1$ with homogeneous coordinates $[\alpha_j,\beta_j]$ and local
affine coordinates $\xi_j:=\alpha_j/\beta_j$ (where $j=1,2$).  Then
${\hat X }$ is realized as the codimension three subspace in
$\P^1[\alpha_1,\beta_1]\times \P^1[\alpha_2,\beta_2]\times
\C^4[z,u,x,y]$ cut by the equations: \bea
\label{hatX}
\beta_1 (u-t_0(z))&=&\alpha_1 x\nn\\
\alpha_2(u-t_1(z))&=&\beta_2 y\\
\alpha_1 \beta_2 (u-t_2(z))&=&\beta_1\alpha_2~~,\nn\\
(u-t_0(z))(u-t_1(z))(u-t_2(z))&=&xy~~.\nn 
\eea The map $\tau$ which
forgets the coordinates on the two $\P^1$ factors implements the
resolution.  The exceptional $\P^1$'s are simply those fibers of
$\tau$ which sit above the double points of $X_0$. One easily checks
that $D^{(1)}_j$ is the factor $\P^1[\alpha_1,\beta_1]$ sitting above
the double point determined by $z^{(1)}_j$, $D^{(2)}_j$ is the factor
$\P^1[\alpha_2,\beta_2]$ sitting above the double point of $X_0$
determined by $z^{(2)}_j$ and ${\tilde D}_j$ is a `diagonal' $\P^1$ in
$\P^1[\alpha_1,\beta_1]\times \P^1[\alpha_2,\beta_2]$, which sits
above the double point determined by ${\tilde z}_j$. More precisely,
the exceptional curves are given by the equations: \bea
\label{exc_curves}
D_j^{(1)}&:&~x=y=0,~z=z_j^{(1)},~u=-W'_2(z_j^{(1)})/3,~\xi_2=0\nn\\
D_j^{(2)}&:&~x=y=0,~z=z_j^{(2)},~u=+W'_1(z_j^{(2)})/3,~\xi_1=\infty\\
{\tilde D}_j~~&:&~x=y=0,~z={\tilde z}~,~~~u=-W'_1({\tilde
  z}_j)/3,~~~\xi_2=-W'_1({\tilde z}_j) \xi_1~~.\nn \eea Note that we
can use $\xi:=\xi_1$ as local affine coordinate on ${\tilde D}_j$
(remember that $W'_1({\tilde z}_j)$ does not vanish, due to our
genericity assumption).

As explained in \cite{OT} (and recalled below), the Hanany-Witten
construction arises upon performing T-duality with respect to the
following $U(1)$ action on ${\hat X}$, which we denote by
${\hat \rho} $: \be
\label{rho_global}
([\alpha_1,\beta_1], [\alpha_2,\beta_2],z,u,x,y)\stackrel{{\hat \rho}(\theta)}{\longrightarrow}
([e^{-i\theta}\alpha_1,\beta_1],
[\alpha_2,e^{i\theta}\beta_2],z,u,e^{i\theta}x,e^{-i\theta} y)~~.  \ee
This projects as follows on the singular space $X_0$: \be
\label{U1proj}
(z,~u,~x,~y)\stackrel{\rho_0(\theta)}{\longrightarrow} (z,~u,~e^{i\theta}x,~e^{-i\theta}y)~~.
\ee The fixed point set of the projected action $\rho_0$ coincides with the
3-section $\Sigma_0$ given in (\ref{Sigma_0}), while the fixed point
locus of (\ref{rho_global}) is its proper transform.  The latter has
three {\em disjoint} components ${\hat C}_j$ which are the proper
transforms of the components $C_j$ of $\Sigma_0$ (in particular
$\tau({\hat C}_j)=C_j$): \bea
\label{hatC}
&&{\hat C}_0:~~x=y=u-t_0(z)=0,~~\alpha_1=\alpha_2=0~~\nn\\
&&{\hat C}_1:~~x=y=u-t_1(z)=0,~~\beta_1=\beta_2=0\\
&&{\hat C}_2:~~x=y=u-t_2(z)=0,~~\beta_1=\alpha_2=0\nn \eea It is easy
to see that $D_j^{(1)}$ touches each of ${\hat C}_0$ and ${\hat C}_2$
at a single point, while $D_j^{(2)}$ has the same behavior with
respect to ${\hat C}_1$ and ${\hat C}_2$. Finally, ${\tilde D}_j$
touches each of ${\hat C}_0$ and ${\hat C}_1$ at a single point.

\subsubsection{Local description and relation to the Hanany-Witten
  construction}

As recalled in the introduction, the $A_2$ quiver theory can also be
obtained through a Hanany-Witten construction which involves a flat
IIA background containing three types of stacks of $D4$-branes stretching between
three NS5-branes.  The relation of this construction to the
geometric engineering given above is implemented by T-duality, as
discussed in a more general context in \cite{OT}.  To see this
explicitly, one considers a local model ${\tilde X}\subset {\hat X}$
of the resolution, which is obtained by gluing three copies
$U_j$ ($j=0\dots 2$) of $\C^3$ (with affine coordinates $x_j, u_j,
z_j$) according to the identifications: \be (x_1,~u_1,
~z_1)=(\frac{1}{u_0}, ~x_0 u_0^2-W'_1(z_0)u_0, ~z_0)~~ \ee and: \be
(x_2, ~u_2, ~z_2)=(\frac{1}{u_1}, ~x_1 u_1^2-W'_2(z_1)u_1, ~z_1)~~.
\ee Then the restricted projection $\tau:{\tilde X}\rightarrow X$ is
given by: \bea
(z;u;x;y)&=&(z_0;~~x_0u_0+t_0(z_0);~~x_0;~~u_0[x_0u_0-W'_1(z_0)][x_0u_0-W'_1(z_0)-W'_2(z_0)])\nn\\
(z;u;x;y)&=&(z_1;~~x_1u_1+t_2(z_1);~~x_1[x_1u_1+W'_1(z_1)];~~u_1[x_1u_1-W'_2(z_1)])
~~~~~~~~~~~~~~~~~~\\
(z;u;x;y)&=&(z_2;~~x_2u_2+t_1(z_2);~~x_2[x_2u_2+W'_2(z_2)][x_2u_2+W'_1(z_2)+W'_2(z_2)];~~u_2)\nn
\eea In this presentation, it is easy to describe only the exceptional
curves $D_j^{(\alpha)}$.  Namely $D_j^{(1)}$ is given by the equations
$z_1=z_j^{(1)}~,~u_1=0$, while $D^{(2)}_j$ is given by
$z_1=z_j^{(2)}~,~x_1=0$.

The Hanany-Witten construction is recovered \cite{OT} by T-duality
with respect to the $U(1)$ action (\ref{rho_global}), which has the
following form in local coordinates: \be
\label{U1action}
(z_j,~u_j,~x_j)\stackrel{{\hat \rho}(\theta)}{\longrightarrow}
(z_j,~e^{-i\theta}u_j,~e^{i\theta}x_j)~~.  \ee The fixed point set of
this action is given by the curves ${\hat C}_j$, whose local equations
are: \be {\hat C}_0:~~u_0=x_0=0~~,~~{\hat C}_1:~~u_2=x_2=0~~,~~{\hat
  C}_2:~~u_1=x_1=0~~.  \ee The action (\ref{U1action}) clearly
stabilizes the exceptional curves $D_j^{(\alpha)}$.  Under T-duality,
the loci ${\hat C}_j$ become three NS5 branes denoted ${\cal N}_j$,
while the D5-branes wrapping $D_j^{(\alpha)}$ are mapped into two
stacks of D4 branes, denoted ${\cal D}_j^{(\alpha)}$. With our
indexing, ${\cal N}_2$ is the central five-brane, while ${\cal N}_0$
and ${\cal N}_1$ are the `outer' five-branes. The intersections
discussed after equations (\ref{hatC}) show that ${\cal D}_j^{(1)}$
stretch between ${\cal N}_2$ and ${\cal N}_0$, while ${\cal
  D}_j^{(2)}$ stretch between ${\cal N}_2$ and ${\cal N}_1$. The
D5-branes wrapping ${\tilde D}_j$ are mapped into $D4$-branes
${\tilde {\cal D}}_j$ stretching between ${\cal N}_0$ and ${\cal N}_1$
(this is possible because the outer NS5 branes are curved and
tilted).  This recovers the Hanany-Witten construction of the $A_2$
quiver theory.

It will be important for our purpose to know the explicit relation
between the flat space coordinates of the T-dual type IIA description
and the complex coordinates $x_1,u_1,z_1$ used in the type IIB
formulation (the relation to the other coordinates $x_j,u_j,z_j$ follows trivially
by gluing). Since we don't know the explicit metric on the minimal
resolution, we cannot identify the metric data, but we can us a
trick\footnote{
It is non-trivial to give a full justification of the identification
(\ref{base_coords})
used below. 
This is because in the presence of a superpotential the metric on the 
resolution need not be compatible with the hyperkahler structure we
introduce for $x_1, u_1$. However, the situation is somewhat similar
to that of \cite{aw} and we expect that an argument along the lines of
that paper can be applied to our situation. Another approach would be
to apply the Buscher formulas \cite{buscher} to a
appropriate supergravity solution for intersecting NS branes and D4 branes.  
In any case, the coordinate identification (\ref{base_coords}) does
reproduce the expected physics. } to determine a good set of
coordinates up to scale  factors.

For this, let us combine $x_1$ and $u_1$ into a quaternion coordinate
$X:=x_1+{\bf j}u_1$, where ${\bf j}$ is the second quaternion
imaginary unit. Then (\ref{U1action}) becomes the standard $U(1)$
action: \be X\longrightarrow e^{i\theta}X~~.  \ee The associated
hyperkahler moment map ${\vec \mu}:\C^2[x_1,u_1]\longrightarrow \R^3$
gives a fibration of $\C^2[x_1,u_1]$ over $\R^3$ whose generic fiber
is a circle (the fiber collapses to a point precisely above the origin
of $\R^3$). We shall denote the Cartesian coordinates of the $\R^3$
base by $x^4,x^5,x^6$ and let $x^7$ be the (periodic) coordinate along
the $S^1$ fiber. Then $x^4+ix^5$ gives the complex part of the
hyperkahler moment map, while $x^6$ is its real part: \be
\label{base_coords}
x^4+ix^5=x_1u_1~~~,~~x^6=\frac{1}{2}(|x_1|^2-|u_1|^2)~~.  \ee The dual
type IIA description is realized in the Minkowski space $\R^{1,9}$
with coordinates $x^0\dots x^9$, where the `internal' coordinates
$x^4\dots x^9$ are related to $x_1,u_1,z_1$ through equation
(\ref{base_coords}) and: \be
\label{extra_coords}
x^8+ix^9=z~~.  \ee It is now easy to see that the NS5-brane
worldvolumes extend along the directions $x^0\dots x^3$ and $x^8,
x^9$, while being localized at $x^7=0$.  The central NS5-brane
${\cal N}_2$ is located at $x^4=x^5=x^6=x^7=0$ and extends in the
directions $x^8$ and $x^9$. The other two NS5-branes sit at
$x^6=\pm \infty, x^7=0$ and are curved in the $x^4, x^5, x^8, x^9$
directions according to the equations: 
\bea
\label{base_coords1}
x^4+ix^5=-W'_1(z)~~,~~x^6=+\infty~~{\rm for}~~{\cal N}_0~~ \eea and:
\be x^4+ix^5=+W'_2(z)~~,~~x^6=-\infty~~{\rm for}~~{\cal N}_1~~.  \ee
The $D4$-brane worldvolumes of ${\cal D}_j^{(\alpha)}$ extend along
$x^0\dots x^3$ as well as $x^6$ and are
localized\footnote{Localization in the direction $x^7$ is due to the
  fact that there is no $B_{NS}$ flux through the exceptional $\P^1$'s.} at $x^4=x^5=x^7=0$
and $z=z_j^{(\alpha)}$.

\subsection{Adding the orientifold }

\subsubsection{The IIB description}

In the type IIB set-up, we consider the case when
$W_1(z)=W_2(-z):=W(z)$. Then $W'_1(z)=-W'_2(-z)=W'(z)$ and we have:
\bea
\label{tspec}
t_0(z)&=&\frac{-2W'(z)+W'(-z)}{3}:=t(z)\nn\\
t_1(z)&=&\frac{-2W'(-z)+W'(z)}{3}=t(-z)\\
t_2(z)&=&\frac{W'(z)+W'(-z)}{3}=-t(z)-t(-z)\nn~~.  \eea In this
situation, we can index the points $z^{(\alpha)}_j$ such that
$z^{(1)}_j=-z^{(2)}_j:=z^+_j$, and we let $z^-_j:=z^{(2)}_j$.  Then
$z^-_j=-z^+_j$ and $z^+_j$ are the roots of $W'(z)$. We also index the
points ${\tilde z}_j$ (which are the roots of the polynomial
$W'(z)-W'(-z)$ ) by positive, zero and negative integers $j$ such that
${\tilde z}_{-j}=-{\tilde z}_j$ (and in particular ${\tilde z}_0=0$).
With these conventions, the exceptional curves $D^{(1)}_j$ and
$D^{(2)}_j$ will be denoted by $D^+_j$ and $D^-_j$; these are
distributed in symmetric pairs with respect to the origin of the
$z$-plane. The curves ${\tilde D}_j$ are also distributed in symmetric
pairs $({\tilde D}_j, {\tilde D}_{-j})$, except for the central curve
${\tilde D}_0$ sitting above $z=0$.  By our genericity assumption,
each of these exceptional curves sits in a distinct fiber of ${\hat
  X}$ over the $z$-plane.

When (\ref{tspec}) are satisfied, the resolution ${\hat X}$
admits a $\Z_2$ symmetry ${\hat \kappa}$ given by:
\be
\label{or_global}
([\alpha_1,\beta_1], [\alpha_2,\beta_2],z,u,x,y)\stackrel{\hat \kappa}{\longrightarrow}
([-\beta_2,\alpha_2], [-\beta_1,\alpha_1],-z,u,-y,-x) \ee which acts
as follows on the affine coordinates $\xi_j=\alpha_j/\beta_j$ of the
two $\P^1$ factors: \be \xi_1\longleftrightarrow -1/\xi_2~~ \ee and
projects to the following involution $\kappa_0$ of $X_0$: \be
\label{or_projected}
(z,~x,~y,~u)\stackrel{\kappa_0}{\longrightarrow} (-z,~-y,~-x,~u)~~.  \ee The action
(\ref{or_global}) stabilizes the central fiber ${\hat X}(0)$, while
(\ref{or_projected}) stabilizes $X_0(0)$.  Tracing through the
equations, we find that the symmetry interchanges $D_j^+$ and $D_j^-$,
while mapping ${\tilde D}_j$ into ${\tilde D}_{-j}$. In particular,
the $\Z_2$ action stabilizes ${\tilde D}_0$, on which it acts as: \be
\label{xi_action}
\xi\longrightarrow \frac{1}{W'(0)}\xi^{-1}~~.  \ee This fixes two
points $p_\pm$ of ${\tilde D}_0$, given by the roots $\xi_{\pm}=\pm
W'(0)^{-1/2}$.

The action (\ref{or_global}) maps ${\hat C}_0$ into ${\hat C}_1$ and
stabilizes ${\hat C}_2$ while fixing the following locus in ${\hat
  X}$: \be
\label{hatO}
{\hat O}:~~y=-x,~~z=0,~~\xi_1\xi_2=-1~~. \ee 
Choosing $\xi_2$, $u$ and $x$ as local coordinates, ${\hat O}$ can be
described by the equations:
\bea
\xi_2 (u+\frac{W'(0)}{3}) &=& -x \nn\\
u-\frac{2W'(0)}{3} &=& - \xi_2^2 
\eea 
This is a smooth
rational curve $x=\xi_2(\xi_2^2-W'(0))$ (parameterized by $\xi_2$)
which sits in the fiber ${\hat X}(0)$ above the point
$z=0$. Its projection is the fixed point locus of the
action (\ref{or_projected}): 
\be
\label{Oplane}
O_0:~~z=0~~,~~x=-y,~~x^2+(u+\frac{W'(0)}{3})^2(u-\frac{2W'(0)}{3})=0~~,
\ee which is a nodal curve sitting in $X_0(0)$. In fact, ${\hat O}$ is
the proper transform of $O_0$ under the blow-up ${\hat
  X}(0)\rightarrow X_0(0)$ (the singular point $(x,u) = (0,-\frac{W'(0)}{3})$
is replaced by $(x,u,\xi_2)=(0,-\frac{W'(0)}{3},\pm W'(0)^{1/2} )$ ).

We shall use ${\hat \kappa}$ as an orientifold action on our type IIB
theory.  Thus our background will contain an O5-`plane`, whose
worldvolume spans the directions $x^0\dots x^3$ and the rational curve
(\ref{hatO}).

\subsubsection{Relation to the orientifolded brane construction in IIA}

\begin{figure}[hbtp]
\begin{center}
  \scalebox{0.7}{\input{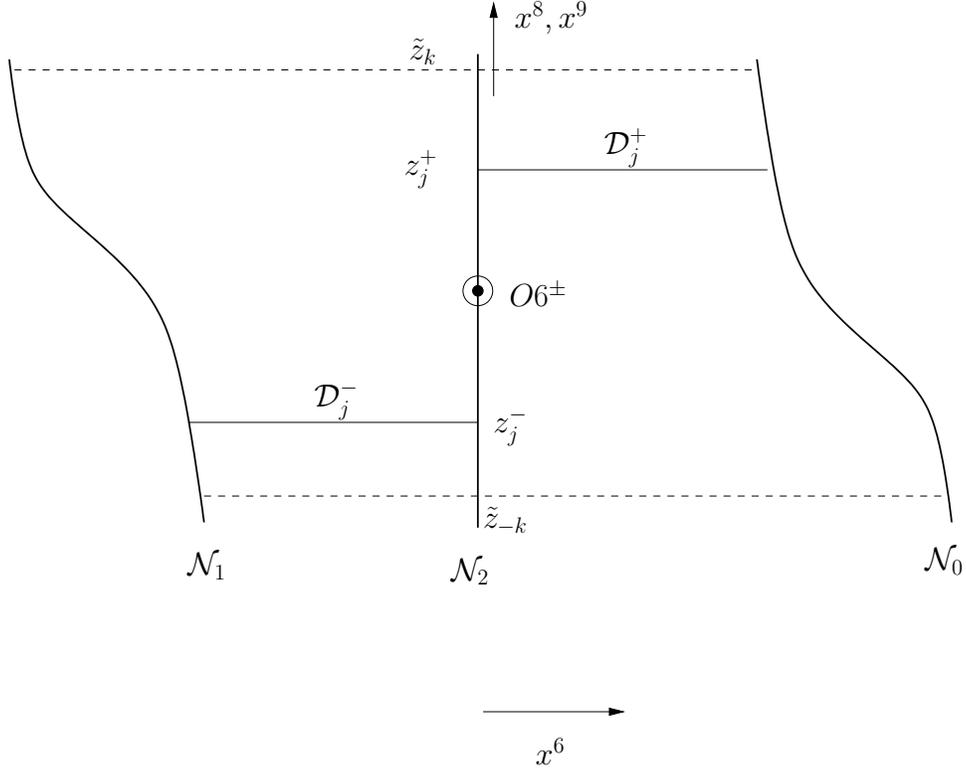}}
\end{center}
\caption{The T-dual brane configuration in IIA on flat $\R^{1,9}$. We show only the coordinates $x^6,x^8$ and $x^9$.
  The coordinates $x^4$ and $x^5$ point outside of the figure. }
\label{brane_config}
\end{figure}

To understand the T-dual orientifold, we start with the local
description of the involution ${\hat \kappa}$ in the patches $U_j$:
\bea
(z_1,~x_1,~u_1)&\stackrel{\hat \kappa}{\longrightarrow}& (-z_1, ~-u_1,~-x_1)~~\nn\\
(z_0,~x_0,~u_0)&\stackrel{\hat \kappa}{\longrightarrow}& (-z_2,~-u_2,~-x_2)~~, \eea which
gives the following local description of the orientifold `plane`: \be
\label{or_plane}
{\hat O}: z=0~~,~~x_1=-u_1~~.  \ee We next note the relation: \be
{\hat \rho} (\theta)\circ {\hat \kappa}={\hat \kappa}\circ {\hat \rho} ({-\theta})~~, \ee which shows
that the IIB involution inverts the $S^1$ coordinate $x^7$. By
T-duality, the orientifold 5-`plane` must thus become an orientifold
6-plane, whose equations are easily extracted from the coordinate
transformations (\ref{base_coords}) and (\ref{extra_coords}): \bea
x^6=x^8=x^9=0~~.  \eea Thus the orientifold extends in the directions
$x^4, x^5$ and $x^7$ and in particular it intersects orthogonally the central
NS5-brane ${\cal N}_2$.  The IIA involution inverts the sign of
$x^6, x^8$ and $x^9$, while leaving the other coordinates unchanged
(see figure \ref{brane_config}). This is precisely the situation
considered in \cite{Karl_nonchiral0, Karl_nonchiral1}\footnote{ Except
  that we do not add any 6-branes in our case. Our IIA coordinates are
  related to those of \cite{Karl_nonchiral1} by the relabeling
  $(x^4,x^5)\longleftrightarrow (x^8, x^9)$.}. The IIA orientifold
permutes the the stacks ${\cal D}_j^+$ and ${\cal D}_j^-$ and the outer NS5 branes ${\cal
  N}_0$ and ${\cal N}_1$. It stabilizes the central five-brane ${\cal
  N}_2$ while acting nontrivially on its worldvolume. We stress that
the O6-plane is orthogonal to the central NS5 brane as well as to
the D4-branes ${\cal D}_j^+$ and ${\cal D}_j^-$. This is quite
different from the situation considered in
the papers \cite{Karl_chiral,bhkl,egkt}, which discussed an alternate orientifold
construction of the same IIA brane configuration. The latter
construction involves an O6-plane which {\em contains} the central
NS5 brane and leads to a {\em chiral} theory containing both
symmetric and antisymmetric matter as well as eight fundamentals. The geometric engineering and
matrix model relevant for that situation are studied in \cite{chiral}.

\subsection{Geometric interpretation of the moduli space}

The brane configurations and their T-dual geometries encode
information about the moduli space of our field theories. A general
discussion of deformations for ${\cal N} = 2, A_k$ quiver field
theories was developed in \cite{OT}.

Consider first the configuration without the orientifold. In ${\cal
  N}=1$ language, the ${\cal N} = 2, U(N_1) \times U(N_2)$ theory with
2 stacks of $N_1$ and $N_2$ D5 branes wrapped on the two exceptional
$\P^1$'s of the resolution of an $A_2$ singularity has the $A_2$
quiver superpotential: \be \label{quiver_superpot}
W=\tr_2(Q_{12}\Phi_1Q_{21}) - \tr_1(Q_{21}\Phi_2Q_{12})~~ \ee where
$\Phi_j$ are the adjoint chiral multiplets obtained by decomposing the
${\cal N}=2$ vector multiplets and $Q_{12}$ and $Q_{21}$ are chiral
multiplet bifundamentals in the representations $({\bar N_1}, N_2)$
and $(N_1, {\bar N}_2)$ respectively.  This is deformed to an ${\cal
  N}=1$ quiver field theory by adding superpotentials $\tr_1 W_1(\Phi_1)$
and $\tr_2 W_2(\Phi_2)$, which corresponds to fibering the $A_2$
singularity over the $z$-plane.  Here $W_\alpha$ are polynomials of
degrees $d_\alpha$ and their derivatives determine the fibration
data as in (\ref{X0}), (\ref{A2_fibration}).

In order to determine the ${\cal N} = 1$ moduli space, one solves the
F-term equations and divides by the complexified gauge group
\cite{Cachazo_Vafa}.  The generic vacuum arises by taking
$\Phi_\alpha={\rm diag}(z_1^{(\alpha)} \Id_{N_1^{(\alpha)}}\dots
z_{d_\alpha}^{(\alpha)}\Id_{N_{d_\alpha}^{(\alpha)}}, {\tilde z}_1\Id_{{\tilde
    N}_1}\dots {\tilde z}_k\Id_{{\tilde N}_{\tilde d}})$ where
$z_j^{(\alpha)}$ are roots of $W'_\alpha$, ${\tilde z}_l$ are the
roots of $W'_1(z)+W'_2(z)$ (${\tilde d}$ is the degree of this latter
polynomial) and we have $\sum_{j}{N_j^{(\alpha)}}+\sum_{l}{{\tilde
    N}_l}=N_\alpha$.  This leads to the residual gauge group: \bea
\label{nu1}
\prod_{j} U(N^{(1)}_j) \times \prod_{j} U(N^{(2)}_j) \times \prod_{l}
U({\tilde N}_l) \eea where $U(N^{(\alpha)}_j)$ are embedded in
$U(N_\alpha)$ (corresponding to zero eigenvalues for the
bifundamentals) and $U({\tilde N}_l)$ are diagonally embedded in
$U(N_1) \times U(N_2)$ (corresponding to non-zero eigenvalues for the
bifundamentals).

In terms of the geometry discussed above, this corresponds to
$N_j^{(\alpha)}$ D5 branes wrapped on the curves $D^{(\alpha)}_j$ and
${\tilde N}_l$ D5 branes on the curves $\tilde{D}_l$. The T-dual brane
configuration contains several stacks of D4 branes. The stacks ${\cal
  D}_j^{(1)}$ (respectively ${\cal D}_j^{(2)}$) contain $N^{(1)}_j$
(respectively $N^{(2)}_j$) D4 branes and stretch from the central
NS5 brane ${\cal N}_2$ to the left (respectively right)
NS5 branes ${\cal N}_0$ and ${\cal N}_1$. The stacks ${\tilde {\cal
    D}}_j$ contain ${\tilde N}_l$ D4 branes and stretch from the left
NS5 brane ${\cal N}_0$ to the right NS5 brane ${\cal N}_1$.

What happens if one adds an orientifold plane? There are two types of
orientifolds in the IIA construction, either an O4 plane parallel to
the D4 branes or an O6 plane which is orthogonal to the D4 branes.  In
the T-dual geometry, the O4 plane becomes an O5 'plane' wrapped on one of the
$\P^1$ cycles (this is the configuration used in \cite{ashok}). On the
other hand, the O6 plane can be chosen in at least two different ways,
which were discussed in \cite{Karl_nonchiral0, Karl_nonchiral1} and
\cite{Karl_chiral,bhkl,egkt} respectively. The first choice
\cite{Karl_nonchiral0, Karl_nonchiral1} is to take the O6 plane to be
orthogonal not only to the D4 branes but also to the central NS5
brane.  This leads to the non-chiral theories studied in the present
paper.  With the second choice \cite{Karl_chiral,bhkl,egkt}, the O6 plane is
orthogonal to the D4-branes but contains the central NS5 brane,
which leads to the chiral theories studied in \cite{chiral}.

With the first choice of O6 plane, the dual IIB orientifold has the
action discussed above (see also \cite{sv}, \cite{eot} for related
though simpler models).  As we saw in the previous subsection, the
orientifold symmetry requires $W_1(z)=W_2(-z):=W(z)$, in which case 
the polynomial $W'_1(z)+W'_2(-z)=W'(z)-W'(-z)$ has degree 
$2\delta+1$, where $\delta=\left[\frac{d-1}{2}\right]$ (see \cite{us}). 
The orientifold projection forces a symmetric distribution of the exceptional curves
$D^{(1)}_j$ and $D^{(2)}_j$ (now denoted by $D_j^+$ and $D_j^-$) and
an arrangement of $\tilde{D}_l$ into symmetric pairs $(\tilde{D}_l,
\tilde{D}_{-l})$, together with the central curve $ \tilde{D}_{0}$.
In the IIA brane construction, this means that we symmetrically
identify the $N^{(1)}_j$ D4 branes ${\cal D}^{(1)}_j={\cal
  D}_j^+$ with the $N^{(2)}_j$ D4 branes ${\cal
  D}_j^{(2)}={\cal D}_j^-$. We also identify the D4 branes
${\tilde D}_j$ with ${\tilde D}_{-j}$ which go
from the left NS5 brane to the right NS5 brane and are located at
opposite positions along the middle NS5 brane, except for the stack
${\tilde {\cal D}}_0$ of D4 branes located at $z = 0$, which
intersects the O6 plane and is mapped to itself under the orientifold
action. For these identifications, one enumerates $z_j^{(\alpha)}$ and
${\tilde z}_j$ such that $z_j^{(1)}=z_j^+=-z_j^-=-z_j^{(2)}$ and
${\tilde z}_{-l}=-{\tilde z}_l$. One also takes $N_1=N_2=N$ as well as
$N_j^{(1)}=N_j^{(2)}$ and ${\tilde N}_{-l}={\tilde N}_l$. 

As explained in \cite{Karl_nonchiral0, Karl_nonchiral1} and
\cite{us}, the orientifold projection on the bifundamental fields
produces a symmetric or antisymmetric field and its conjugate (the
symmetric field appears if one uses an $O6^+$ plane and the
antisymmetric field appears for an $O6^-$ plane).  Then, the
identified $N^{(1)}_j=N_j^{(2)}$ D4 branes ${\cal D}_j^{(1)}\equiv {\cal
  D}_j^{(2)}$ correspond to zero vev for the symmetric/antisymmetric
field and nonzero vev for the adjoint field. Since the
superpotentials $W_j(\Phi_j)$ correspond to tilting and bending of the
outer NS5-branes, these D4 branes are displaced along the middle
NS5 brane ${\cal N}_2$ and they do not intersect the O6 plane.

The stacks of identified $\tilde{N}_l={\tilde N}_{-l}$ D4 branes
${\tilde {\cal D}}_l\equiv {\tilde {\cal D}}_{-l}$ correspond to
nonzero vevs for both the adjoint and symmetric/antisymmetric
fields.  Such D4 branes are displaced with
respect to the central NS5 brane and the O6 plane and do not touch any of
them, stretching directly between the left and right NS5 branes.

The last stack of D4 branes ${\tilde {\cal D}}_0$ corresponds to zero
vev for the adjoint field but nonzero vev for the symmetric/antisymmetric field.
Such D4 branes are displaced along the O6 plane and touch the left and
right NS5 branes, but not the middle NS-brane ${\cal N}_2$. Because
these D4 branes touch the O6 plane, the projected gauge group is
$SO(N_0)$ or $Sp(N_0/2)$, depending on whether $s=+1$ or $s=-1$.

This discussion agrees with the results of Section 2.1 of \cite{us}
and recovers the fact that in the generic supersymmetric vacuum
the adjoint field has a vev:
\be
\label{phi_vev_sa}
\Phi={\rm diag}(0_{{\tilde N}_0}, z_1^+ 1_{N_1} \dots z_d^+ 1_{N_d}, 
{\tilde z}_1 1_{{\tilde N}_1}, -{\tilde z}_1 1_{{\tilde N}_1} ~\dots ~{\tilde z}_\delta 
1_{{\tilde N}_\delta}, -{\tilde z}_\delta 1_{{\tilde N}_\delta} )~~,
\ee
while the residual gauge group is\footnote{We use the convention that
  $U(0)$ is the trivial group.}: \bea \prod_{i=1}^n U(N_i) \times
\prod_{j=1}^{n} U(\tilde{N}_j) \times G_0 \eea where
$\sum_{i=1}^{d} N_i + 2 \sum_{l=1}^{\delta} \tilde{N}_l + {\tilde N}_0 = N$
with $\delta=\left[\frac{d-1}{2}\right]$ and $G_0=SO({\tilde N}_0)$ if $s=+1$, respectively
$G_0=Sp({\tilde N}_0/2)$ if $s=-1$.

\section{The geometric transition and the effective superpotential}
\label{transition}

In this section we consider the geometric transition of \cite{vafa, civ, Cachazo_Vafa, Cachazo_Vafa_more},
which replaces the D5 branes by fluxes through the 3- cycles of a
deformed geometry. The D5 branes wrapped on the $\P^1$ cycles go
through the transition in the usual way \cite{vafa,civ,Cachazo_Vafa, Cachazo_Vafa_more},
so it suffices to concentrate on understanding the orientifold
projection after the transition.

\subsection{The orientifold after the geometric transition}

Let $X$ be the smoothing of the singular threefold $X_0$, which is
described by the equation:
\be
\label{X}
xy=u^3-p(z)u-q(z)~~,
\ee where (as in \cite{Cachazo_Vafa})
we only consider log-normalizable deformations (the explicit
expression of $p,q$ for such deformations is recalled in equation
(\ref{pq}) of Section 4.3.2 below).

After the geometric transition ${\hat X}\longrightarrow
X_0\longrightarrow X$ of \cite{vafa,civ,Cachazo_Vafa, Cachazo_Vafa_more},
the D5-branes wrapping the exceptional fibers of ${\hat X}$ will be replaced by fluxes
through the $S^3$ cycles created by smoothing. If one starts with a
$\Z_2$ symmetric hypersurface $X_0$ (which is achieved by requiring
$W_1(z)=W_2(-z):=W(z)$), then one can restrict the smoothing $X$ by
requiring that $p$ and $q$ are even, so that $X$ will admit the $\Z_2$
symmetry (\ref{or_projected}). In this situation, we add the
orientifold and ask what happens after the transition.

Since the deformed Calabi-Yau admits the action (\ref{or_projected}),
it is clear that the orientifold will survive the transition, being
mapped into an orientifold of IIB with geometric action $\kappa$ given
by equations (\ref{or_projected}).  The $A_2$ fibration $X$ admits a
multisection $\Sigma$ (the deformation of $\Sigma_0$) given by the
equations $x=y=0$, which imply:
\be
\label{Sigma}
\Sigma:~~u^3-p(z)u-q(z)=0~~.  \ee Let us write this Riemann surface as a triple
cover of the $z$-plane: \be
u^3-p(z)u-q(z)=(u-u_0(z))(u-u_1(z))(u-u_2(z))~~ \ee where
$u_2(z)=-u_0(z)-u_1(z)$ and we index the branches $u_j(z)$ such that
they are deformations of $t_j(z)$. Then $\Sigma$ is invariant under
the involution and we have $u_0(-z)=u_1(z)$ and $u_2(-z)=u_2(z)$.
The smoothing replaces the double points 
$z_i^\pm$ and ${\tilde z}_j$ of $\Sigma_0$ with cuts of $\Sigma$ which 
we denote by $I_{\pm i}$ and ${\tilde I}_j$. The cuts $I_i$ connect the
branches $u_0$ and $u_2$ if $i>0$ and the branches $u_1$ and $u_2$ if
$i<0$, while each of the cuts ${\tilde I}_j$ connects the branches
$u_0$ and $u_1$. 
As discussed in \cite{us}, 
these cuts are distributed symmetrically with respect to the origin of
the $z$-plane, i.e. we have $I_{-i}=-I_i$ and ${\tilde I}_{-j}=
-{\tilde I}_j$. In particular, the central cut ${\tilde I}_0$ is
symmetric under the point reflection $z\rightarrow -z$.

The orientifold action (\ref{or_projected}) on $X$ has fixed locus
given by:
\be
\label{def0}
O:~~z=0~~,~x=-y~~,
\ee
which defines a (generically smooth)
elliptic curve lying inside the central fiber $X(0)$:
\be
\label{defOplane}
x^2+u^3-p(0)u-q(0)=0~~.
\ee This locus corresponds to an O5-`plane'
which survives the geometric transition.

Thus we end up with a compactification with NS-NS and R-R fluxes plus
an orientifold five-`plane' whose internal part is given by (\ref{defOplane}).  As explained
in \cite{Vafa_or}, such a compactification will produce a
superpotential which receives contributions from fluxes and from the
orientifold fixed locus. Since the geometric transition corresponds to 
confinement in the low energy field theory on the noncompact part of the
D5-brane worldvolumes, the flux-orientifold superpotential after the
transition can be identified with the effective superpotential of this
theory for the glueball superfields.

\subsection{The matrix model prediction for the flux-orientifold superpotential}
\label{comp_matrix}

In order to compute the effective superpotential from geometry, we must calculate the 
periods of the holomorphic three-form of the deformed Calabi-Yau $X$.
For this, recall from \cite{civ, Cachazo_Vafa, Cachazo_Vafa_more} that
every one-cycle of the Riemann surface $\Sigma$ defines a 3-cycle of
$X$ obtained by considering a certain $S^2$ fibration associated with that one-cycle.
Accordingly, we consider the following set of three-cycles of
$X$. Let $A_i$ be a three-cycle which is obtained as an $S^2$
fibration over a one-cycle $a_i$ of $\Sigma$
which surrounds the cut $I_i$. We can view $A_i$
as the three-cycle produced by smoothing the singular point of $X_0$
siting above $z_i$.
According to our
considerations in the previous section this corresponds to the
$U(N_i)$ component of the unbroken gauge group.  We choose $A_i$ such that the
$\Z_2$ symmetry $\kappa$
maps it into the cycle $A_{-i}$, and we consider the $\kappa$-invariant linear combination $A_i+A_{-i}$ for $i=1\dots
d$. Similarly, we consider $\kappa$-invariant cycles $\tilde A_j +
\tilde A_{-j}$ for $j=1\dots\delta$, where the 3-cycles ${\tilde
  A}_j$  correspond to one-cycles ${\tilde a}_j$ of $\Sigma$ surrounding 
the branch cut of type ${\tilde I}_j$ and to the component $U(\tilde
N_j)$ of the unbroken gauge group.
They are produced by smoothing the singular point of $X_0$ which
sits above ${\tilde z}_j$, and are chosen such that $\kappa({\tilde
  A}_j)={\tilde A}_{-j}$.
Finally, we consider a three-cycle ${\tilde A}_0$ which is invariant under
$\kappa$. It corresponds to a one-cycle 
${\tilde a}_0$ on $\Sigma$ which surrounds the cut ${\tilde I}_0$ 
and arises by smoothing the singular point of $X_0$ sitting above the
origin.  This 3-cycle also corresponds to the component $G_0=$
$SO({\tilde N}_0)$ or $Sp({\tilde N}_0/2)$ of the unbroken gauge
  group. 

We are thus led to consider two classes of A-periods of the holomorphic 3-form. The first
class is given by: 
\be
\label{periods} 2\pi iS_i = \frac 1 2 \int_{A_i+A_{-i}}
\Omega = \int_{A_i} \Omega\,, \ee where the last equality holds
because of the invariance of $\Omega$ under $\kappa$. 
We also have the periods:
\be\label{tildeperiods} 2\pi i\tilde S_j = \frac 1 2 \int_{\tilde
  A_j+\tilde A_{-j}} \Omega = \int_{\tilde A_j} \Omega\,, \ee and
\be\label{zeroperiod}  2 \pi i \tilde S_0 = \frac 1 2 \int_{\tilde A_0}
\Omega\,.  
\ee 
The cycles which define the periods
$S_i$ for $i=1\dots d$ and $\tilde S_j$ for $j=1\dots \delta$ will be called
{\em long} invariant cycles, whereas $\tilde A_0$ is a {\em short}
invariant cycle
\footnote{The notion of long
and short invariant cycles plays a prominent role in the theory
of boundary singularities \cite{arnold}.}.

The fluxes of the three-form $H$ are:
\bea
\label{ifluxes}
N_i &=&  \int_{A_i} H~~,~~i=-d\dots -1, 1\dots d\nn\\
\tilde N_j &=& \int_{\tilde A_j} H~~,~~j=-\delta \dots \delta~~.
\eea

The collection $A_i, {\tilde A}_j$ can be completed by considering 
a set of 3-cycles $B_i({\Lambda})$ and ${\tilde B}_j({\tilde \Lambda})$
with intersection numbers:
\bea
\langle A_i, B_j \rangle &=&  \delta_{ij},~~\mathrm{for}~~ i,j=-d\dots
-1, 1\dots d\,\\
\langle \tilde A_i, \tilde B_j \rangle &=&
\delta_{ij},~~\mathrm{for}~~ i,j=-\delta \dots \delta ~~\nn
\eea and zero otherwise. 

The B-cycles depend explicitly on two complex cutoffs $\Lambda,
{\tilde \Lambda}$ whose absolute values we take to be large. Then 
$B_i(\Lambda)$ and ${\tilde B}_i({\tilde \Lambda})$ are constructed as certain
$S^2$ fibrations over one-cycles $b_i(\Lambda)$ and ${\tilde
  b}_j({\tilde \Lambda})$ on $\Sigma$ which are constrained to pass through the
points of $\Sigma$ obtained by lifting $\Lambda$ and ${\tilde
  \Lambda}$ from the $z$-plane. Namely, $b_i(\Lambda)$ connects the lifts of $\Lambda$ to the
branches $u_0$ and $u_2$ (if $i>0$) respectively $u_1$ and $u_2$ (if
$i<0$), while ${\tilde b}_j({\tilde \Lambda})$ connects the two lifts of
${\tilde \Lambda}$ to the branches $u_0$ and $u_1$. Mathematically,
this amounts to working with an open-closed (punctured) Riemann surface obtained
by removing the lifts of $\Lambda$ and ${\tilde \Lambda}$ from the
original closed Riemann surface $\Sigma$.  

Thus the remaining periods are given by: 
\bea
\Pi_i &=& \int_{B_i(\Lambda)}{\Omega}~~\mathrm{for}~~i=-d\dots -1, 1\dots d~~,\nn\\
\tilde \Pi_j &=&\int_{{\tilde B}_j({\tilde
  \Lambda})}{\Omega}~~\mathrm{for}~~j=-\delta \dots
\delta~~.
\eea 

Relations similar to (\ref{ifluxes}) hold for the gauge couplings: 
\bea
\alpha_i &=& \int_{B_i(\Lambda)} H
\,,~~\mathrm{for}~~i=1\dots d\,\\
\tilde \alpha_j &=& \int_{{\tilde B}_j({\tilde \Lambda})} H
\,,~~\mathrm{for}~~j=0\dots\delta~~.
\eea

As in \cite{civ, Cachazo_Vafa, Cachazo_Vafa_more}, the period
integrals of $\Omega$
can be reduced to corresponding period integrals of the 
meromorphic one-form $ u \frac{dz}{2\pi i}$ on the Riemann
surface (\ref{Sigma}) (for this, the normalization of
$\Omega$ must be chosen appropriately):
\bea
S_i &=& \int_{a_i}{u \frac{dz}{2\pi i}}~~~~,~~~~{\tilde S}_j = \int_{{\tilde a}_j}{u
  \frac{dz}{2\pi i}}~~~,~~~\tilde S_0 = \frac{1}{2}\int_{{\tilde a}_0}{u \frac{dz}{2\pi i}} \nn\\
\Pi_i &=& \int_{b_i(\Lambda)}{u dz}~~,~~\tilde \Pi_j = 
\int_{{\tilde b}_j({\tilde \Lambda})} {u dz}~~,~~ 
\tilde \Pi_0 = \int_{{\tilde b}_0({\tilde \Lambda})} {u dz} 
~~.
\eea

The flux-orientifold superpotential takes the form: 
\be
\label{Weff}
W_{eff}=\sum_{i=1}^d{\left[N_i\Pi_i+2\pi i\alpha_i S_i\right]}+
\sum_{j=1}^\delta{\left[{\tilde N}_j {\tilde \Pi}_i+{2\pi i\tilde \alpha}_j
    {\tilde S}_j\right]} +\frac{{\tilde N}_0}{2} {\tilde
  \Pi}_0+2\pi i\tilde\alpha_0 \tilde S_0+ 4F_1\,.  \ee

The first terms arise from the fluxes, while the last term is the
orientifold contribution. The factor of $1/2$ in front of ${\tilde
  N}_0 {\tilde \Pi}_0$ is due to the fact that ${\tilde
  N}_0$ as defined in (\ref{ifluxes}) arises by integrating $H$ over
the short cycle ${\tilde A}_0$.  Up to the orientifold contribution $F_1$, this is precisely half
of the flux-superpotential of an $A_2$ quiver theory with a
$\kappa$-symmetric arrangement of fluxes and deformations.
We have the special geometry relations:
\be
\label{sg}
\Pi_i=\frac{\partial F_0}{\partial S_i}~~,~~{\tilde
  \Pi}_j=\frac{\partial F_0}{\partial {\tilde S}_j}~~,
\ee
where $F_0$ is the closed string prepotential. 
We are now going to use matrix model arguments to show that
the orientifold contribution $F_1$ takes the form:
\be
\label{F1geom}
F_1=-\frac{s}{4}\int_{{\tilde b}_0({\tilde \Lambda})}{u\,dz}~~.
\ee

As in \cite{DV}, the low energy effective superpotential of our field theory
is encoded by the dynamics of the topological sector of open strings connecting the B-type branes
wrapping the exceptional $\P^1$ cycles of ${\hat X}$.
In our case, this two dimensional theory is obtained by reducing the holomorphic Chern-Simons
action, similar to the argument given in \cite{DV2} for the $ADE$ quiver theories, but
including the orientifold projection. One easily finds that the
open topological sector reduces to the holomorphic \cite{holo} matrix model constructed in
\cite{us}. This holomorphic matrix model was studied in
\cite{us, holo, Naculich, Naculich1}. As explained in those
references, the matrix model leads to a prescription for computing the
gaugino superpotential in the ${\cal N}=1$ $U(N)$ field theory with
one adjoint and one symmetric or antisymmetric chiral multiplet.  Through geometric
engineering, the latter is realized as the low energy limit of the
flux-orientifold compactification on $X$, and the gaugino
superpotential must coincide with the flux-orientifold superpotential
(\ref{Weff}).

As explained in
\cite{us}, the planar limit of this model's traced resolvent coincides
up to a shift with the sheet $u_0$ of
the Riemann surface $\Sigma$, which is identified with the matrix model's spectral curve.
Following the general prescription of the Dijkgraaf-Vafa correspondence, the closed string prepotential $F_0$
is realized as the planar limit of the matrix model's free energy
($=$ microcanonical generating function), while the orientifold term 
$F_1$ gives
the $\R\P^2$ contributions to the matrix model free energy. In
particular, the orientifold term $F_1$ can be expressed in terms of
matrix model data. It was furthermore proven in \cite{ino} that the
$\R\P^2$ part of the matrix model partition function contributes with
a relative factor of $4$ to the gluino superpotential (\ref{Weff}).
Notice that the D5 brane charge of our orientifold is given by $-s$.
Therefore combining (\ref{F1geom}) and (\ref{Weff}) it turns out that
the orientifold contribution to the superpotential is proportional to
its D5 brane charge.
 
We shall use the results of \cite{us} in order
to extract the geometric description of $F_1$ given in (\ref{F1geom}).

Recall from \cite{us} that the exact loop equations for the traced matrix model resolvent 
$\omega(z) = \tr\left(\frac{1}{z-M}\right)$ (where $M$ is the matrix
associated with $\Phi$) have the form:
 \begin{eqnarray}
\label{ex_loop}
  &&\big\langle \omega(z)^2+\omega(z)\omega(-z)+\omega(-z)^2 \big\rangle
  \;=\; \int_{\gamma} \frac{dx}{2\pi i} \, \frac{2x U'(x)}{z^2{-}x^2}
  \, \big\langle \omega(x)\big\rangle
  \quad , \qquad \nn\\[8pt]
  &&\big\langle \omega(z)^2\omega(-z) + \omega(z)\omega(-z)^2 \big\rangle - \;
  \frac1{N^2}
  \,\frac{\langle\omega(z){+}\omega(-z){-}2\omega(0)\rangle}{4\,z^2}
  \\[4pt]
  && \qquad \;=\; \int_{\gamma} \frac{dx}{2\pi i} \frac{2x
    U'(x)}{z^2{-}x^2} \big\langle \omega(x)\omega(-x)\big\rangle \nn~~.
\end{eqnarray}
where $U(z) = W(z) + (t_{-1} + \frac{s}{2N})\ln(z)$ and $\gamma$ is a
contour encircling the poles of $\omega(z)$ but not the point $z$
nor the poles of $\omega(-z)$.
By expanding these loop equations it was shown in \cite{us} that the
contribution of $\R\P^2$ diagrams to the matrix model free energy is
given by: \be
\label{F1mat}
F_1=\frac{s}{2}\frac{\partial F_0}{\partial t_{-1}}|_{t_{-1}=0}~~,
\ee where $F_0$ is the contribution of $\P^1$ diagrams to the free
energy of a deformed model obtained by adding the logarithmic term
$t_{-1}\ln z$ to the original matrix model potential $W(z)$.  As
explained in \cite{us}, the planar free energy of this deformed model
obeys certain Whitham-like constraints, one of which has the form
\footnote{Here we use an `almost Hermitian formulation' introduced in 
\cite{us} based on the work of \cite{holo}. This amounts
  to requiring the eigenvalues of $M$ to lie on the displaced real
  axis $\R+i\epsilon$, where one takes the regulator $\epsilon$ to
  zero at the very end of all computations. This formulation is
  correct only if $W$ has even degree (since otherwise the matrix
  model partition function diverges). A similar relation can be
  written when $W$ has odd degree, by using the general set-up of
  \cite{holo}.  In that case, one requires the eigenvalues of $M$ to
  lie on a more general contour $\gamma$ in the complex plane, which must be
  chosen such that $\gamma\cap (-\gamma)=\emptyset$. Then one must
  consider the limit when $\gamma$ coincides with $-\gamma$ at the end
of all computations, which corresponds to working with a limiting 
statistical ensemble of holomorphic matrix models \cite{holo}.}: \be
\label{Whitham}
\frac{\partial F_0}{\partial
  t_{-1}}=-\int_{\R+i\epsilon}{d\lambda}{\rho_0(\lambda)\log\lambda}~~.  \ee As
discussed in \cite{us}, adding the logarithmic term $t_{-1}\ln z$
to the matrix model potential has the effect of replacing $\Sigma$
with a singular algebraic curve. The latter can be mapped to a smooth Riemann surface of higher genus,
which has the effect of introducing new cuts for the deformed model.
This means that in the presence of the logarithmic deformation, the eigenvalues of the deformed matrix model
can accumulate not only along $I_i$ and ${\tilde I}_j$, but also along
  certain new loci in the complex plane (a precise description of the
  new cuts is given in Section 3.6 of \cite{us}). Since
equation (\ref{F1mat}) only requires the result of (\ref{Whitham}) for
$t_{-1}=0$, we can in fact neglect these new cuts and evaluate the
right hand side of (\ref{Whitham}) in the {\em undeformed} theory. To
write the result in geometric fashion, let us consider the function:
\be
\label{Psi}
\Psi(\lambda)=\int{d\lambda'\rho_0(\lambda')\left[\ln
    |\lambda+\lambda'|+\ln|\lambda-\lambda'|
  \right]}-W(\lambda)-W(-\lambda)~~, \ee which was introduced in
Section 4 of \cite{us}. Combining (\ref{F1mat}),(\ref{Whitham}) and
(\ref{Psi}), we find: \be F_1=-\frac{s}{4}\Psi(0)-\frac{s}{2}W(0)~~,
\ee where $\Psi(0)$ should be evaluated for $t_{-1}=0$. As shown in
\cite{us}, $\Psi$ is constant along each of the cuts ${\tilde
  I}_j$, where its value coincides with the planar chemical potential
${\tilde \mu}^{(0)}_j$ up to an overall constant which can be fixed by
choosing a cutoff ${\tilde \Lambda}$ at infinity. Upon making this
choice, the chemical potentials ${\tilde \mu}_j^{(0)}$ become equal
with the periods ${\tilde \Pi}_j({\tilde \Lambda})$ computed over the
B-cycles ${\tilde b}_j({\tilde \Lambda})$ discussed at the beginning
of the present section.
In particular, we have ${\tilde \mu}_0={\tilde \Pi}_0({\tilde
  \Lambda})$ and thus: \be F_1=-\frac{s}{4}{\tilde \Pi}_0({\tilde
  \Lambda})-\frac{s}{2}W(0)~~.  \ee We are free to pick the
cutoff ${\tilde \Lambda}$ for the ${\tilde \Pi}$-periods to be
different from the cutoff $\Lambda$ for the $\Pi$-periods.  In
particular, we can pick ${\tilde \Lambda}$ so that we absorb the
contribution $-\frac{s}{2}W(0)$ to $F_1$. With this choice, we recover equation (\ref{F1geom}).

\begin{figure}[hbtp]
\begin{center}
  \scalebox{0.7}{\input{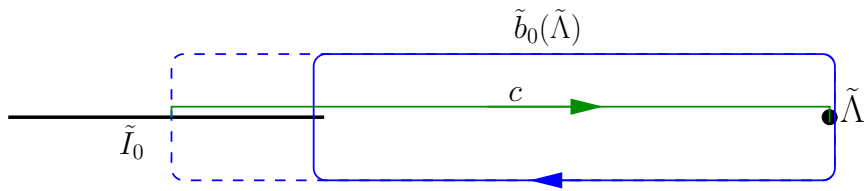}}
\end{center}
\caption{The orientifold contribution to the superpotential is
  proportional to the period ${\tilde \Pi}_0({\tilde \Lambda})$. The
  relevant integral can be expressed as the integral of $u_1-u_0$ over
  the curve $c$ lying in the $z$-plane.}
\label{o5contrib}
\end{figure}

Since ${\tilde \Pi}_0=\frac{\partial F_0}{\partial {\tilde S}_0}$, we have:
\be
\label{F1S0}
F_1=-\frac{s}{4}\frac{\partial F_0}{\partial {\tilde S}_0}~~.
\ee
A similar  relation was observed in \cite{ashok} for the case of SO/Sp gauge theories with
adjoint matter.

\subsection{Comparison with the proposal of \cite{Vafa_or}}

Let us compare relation (\ref{F1geom}) with the proposal of \cite{Vafa_or}.
For this, we write (\ref{F1geom}) in the form:
\be
\label{F1chain}
F_1=-\frac{s}{4}\int_{c}{[u_1(z)-u_0(z)]dz}~~,
\ee
where $c$ is a curve in the $z$-plane which connects the origin with
the point ${\tilde \Lambda}$ (see figure \ref{o5contrib}). To arrive
at (\ref{F1chain}) we used analyticity of $u_0$ and $u_1$ to deform
the projection of ${\tilde b}_0({\tilde \Lambda})$ toward the origin
of the $z$-plane as shown in figure \ref{o5contrib} (this doesn't
change the value of the integral because $u_1(x+i0)=u_0(x-i0)$ along
the cut ${\tilde I}_0$).

To compare with \cite{Vafa_or}, we shall express (\ref{F1chain}) as an
integral of $\Omega$ over a 3-chain ${\cal C}$ in the deformed Calabi-Yau space $X$.
Following the ideas of \cite{civ, Cachazo_Vafa, Cachazo_Vafa_more},
this  3-chain is defined as the total space of an $S^2$ fibration over
the curve $c$. To construct this fibration explicitly, we write the deformed Calabi-Yau (\ref{X}) in the form:
\be
\label{X_alt}
s^2+t^2=(u-u_0(z))(u-u_1(z))(u-u_2(z))~~,
\ee
where $x=s+it$ and $y=s-it$.  Let us fix the point $z$ and consider
the segment $I_{01}(z)=[u_0(z), u_1(z)]$ which connects the points
$u_0(z)$ and $u_1(z)$ in the $u$-plane. Picking a square root
$\sigma(u, z)$ of the right hand side of (\ref{X_alt}), we have
$s^2+t^2=\sigma(u,z)^2$, where $\sigma(u,z)$ vanishes when $u$
coincides with one of $u_j(z)$. Let $C(u,z)$ denote the circle
$\alpha^2+\beta^2=1$ obtained by requiring that
$\alpha:=\frac{s}{\sigma(u,z)}$ and $\beta:=\frac{t}{\sigma(u,z)}$ are
both real. Varying $u$ inside the interval $I_{01}(z)$, we obtain a
two-sphere $S_z^2$ given as an $S^1$ fibration over this segment whose
fiber collapses to a point at the ends of the interval. Finally, we
fiber this $S^2$ over $c$ by
letting $z$ vary along this curve.  This gives the desired 3-chain ${\cal C}$ in the
deformed Calabi-Yau.
The boundary of ${\cal C}$
consists of the two-spheres sitting above ${\tilde \Lambda}$ and above
the origin of the $z$-plane:
\be
\partial {\cal C}=S^2_0 \cup S^2_{\tilde \Lambda}~~.
\ee

This boundary intersects the orientifold plane (\ref{def0}) for $x=-y
\Leftrightarrow s=0$, which gives $\alpha=\frac{s}{\sigma(u,0)}=0$ and
thus $\beta=\pm 1$. This is a circle $\Gamma$ inside $S^2_0$ traced by
two opposite points of the circle $C(z,u)$ when $u$ varies along
$I_{01}(0)$. Thus:
\be
(\partial {\cal C})\cap O =\Gamma~~.
\ee

Using the arguments of \cite{civ, Cachazo_Vafa, Cachazo_Vafa_more},
one can immediately show the relation\footnote{This requires an appropriate normalization of $\Omega$.}:
\be
\label{F1Omega}
F_1=-\frac{s}{4}\int_{{\cal C}}{\Omega}~~,
\ee
which reduces to (\ref{F1chain}) upon performing the integral over the
two-sphere fibers of ${\cal C}$. 

According to the proposal of \cite{Vafa_or} the
superpotential contribution of the orientifold should be given by
integrating the holomorphic 3-form $\Omega$ along a three-chain 
whose boundary consists of the internal part $O$ of the orientifold `plane' and a piece sitting
at infinity (which in our case is represented by 
$S^2_{\tilde \Lambda}$ after introducing the cutoff). 
This is almost exactly what expression (\ref{F1Omega}) does. 
However, in our case
the O5 'plane' is noncompact and the 3-chain ${\cal C}$ which
reproduces the result known from the matrix model intersects the
orientifold fixed locus in $X$ along a {\em circle}.
We attribute this phenomenon to the fact that in our case the internal
part $O$ of the O5 'plane' is noncompact.

\section{Engineering of $SO(N)$ and $Sp(N/2)$ gauge theories with adjoint matter}
\label{sosp}

In this section we study the geometric engineering of ${\cal
  N}=1$ supersymmetric gauge theories with orthogonal or symplectic
gauge group with the help of the orientifold action introduced in the
previous sections. It is well known that
the T-dual Hanany-Witten construction allows the use of either
orientifold four-planes or orientifold six-planes \cite{o4o6} as a means to
engineer $SO/Sp$ gauge theories. In fact the type IIB construction
with orientifold five-'planes' T-dual to orientifold
four-planes has already been studied in \cite{eot} and in
relation to matrix models in \cite{ashok}. Here we consider the case of an
orientifold 6-plane, where both NS branes are rotated and/or deformed
with respect to their ${\cal N} = 2$ position in a manner dictated by
  $W'(z)$.

\subsection{Two geometric engineering constructions}

Let us start with the singular $A_1$ fibration $X_{1,0}$ given by:
\be
X_{1,0}:~~xy=(u-t_0(z))(u-t_1(z))\,,
\ee where $t_0(z) = W'(z)$ and $t_1(z)=- W'(z)$.
This fibration admits the two-section:
\be
\label{Sigma_10}
\Sigma_{1,0}:~~x=y=0,~~(u-W'(z))(u+W'(z))=0~~,
\ee
whose irreducible components are two rational curves.

The resolution ${\hat X}_1$ can be described globally as the complete
intersection:
\bea
\beta ( u - t_0(z)) &=& \alpha x \nn\\
\alpha (u - t_1(z)) &=& \beta y \\
(u-t_0(z))(u-t_1(z)) &=& xy~~\nn
\eea
in the ambient space $\P^1[\alpha,\beta]\times \C^4[z, u, x, y]$.
The exceptional $\P^1$'s of the resolution sit above the singular points of $X_{1,0}$, which are determined by
$x=y=u=0$ and $z=z_j$, where $z_j$ are the roots of $W'$. The resolution admits the $U(1)$ action:
\be
\label{U1red}
([\alpha, \beta], z, u, x, y) \longrightarrow
([e^{-i\theta} \alpha, \beta], z, u,e^{i\theta} x, e^{-i\theta}y),
\ee
and T-duality with respect to its orbits allows one to recover the Hanany-Witten
description.

Let us now add an orientifold. By analogy with the
previous sections, we can use the action:
\be
\label{red_or}
([\alpha,\beta],z,u,x,y) \longrightarrow ([-\beta, \alpha], -z, u, -y,
-x)~~,
\ee
which is a symmetry provided that one takes $W(z)$ to be an {\em even} polynomial, whose degree
we denote by $2n$.
Then one can index the critical points of $W$ by $z_j$ with $j=-(n-1)\dots n-1$, such that
$z_{-j}=-z_j$. In particular, we have the critical point $z_0=0$. We let $D_j$ denote the exceptional
$\P^1$  sitting above $z_j$.
The action (\ref{red_or}) maps $D_j$ into $D_{-j}$
and in particular it stabilizes the central exceptional  curve $D_0$,
on which it acts through antipodal involution. Since $W'(0)=0$, the central fiber of $X_{1,0}$ is
an $A_1$ singularity with equation $xy=u^2$, while the central fiber
of ${\hat X}_1$ is its minimal resolution.

The action (\ref{red_or}) projects to the following involution of $X_{1,0}$:
\be
\label{red_or_proj}
(z,u,x,y) \longrightarrow (-z, u, -y,-x)~~,
\ee
whose fixed point set is given by:
\be\label{Olocus}
O_{1,0}: x=-y,~~z=0,~~x^2+u^2=0~~.
\ee
This is a reducible curve whose two rational components $x=\pm i u$ sit in the
central fiber of $X_{1,0}$. The fixed point
set ${\hat O}_1$ of (\ref{red_or}) is is a disjoint union of two rational curves lying inside
the central fiber of the resolved space (the singular point $(u,x)=(0,0)$ 
of (\ref{Olocus}) is 
replaced by the two points $(x,u,\xi)=(0,0,\pm i)$ where $\xi=\alpha/\beta$).
Thus the orientifold action (\ref{red_or}) determines a (disconnected) orientifold 5-`plane' which we
denote by O5 \footnote{Since O5 has two connected components, it
  should be viewed more properly as {\em two} orientifold fixed `planes'
  whose RR charge adds to that of a single O5 plane. This unusual
  situation is due to the fact that we work with a nontrivial
  geometric background and our orientifold `planes' are curved.}.

Since $W$ is an even polynomial, the resolved space ${\hat X}_1$ admits
another holomorphic involution, which acts as:
\be
([\alpha,\beta],z,u,x,y)
\longrightarrow ([ \alpha, \beta], -z, -u, -x, -y)\,.
\label{ofiveprime}~~.
\ee
This orientifold action was used in previous studies \cite{sv, eot, ashok} to geometrically engineer
the $SO(N)$ and $Sp(N/2)$ theories with one adjoint chiral multiplet. As we shall see below, the same theories
can be engineered by using the action (\ref{red_or}), and we are interested in comparing the two realizations.

The fixed point set ${\hat O}_1'$ of the action (\ref{ofiveprime}) coincides with the exceptional curve $D_0$
in ${\hat X}_1$ which sits above the singular point $x=y=z=u=0$ of $X_{1,0}$.
We will denote the associated orientifold 5-'plane' by O5'.
Notice that O5' coincides with the worldvolume of the stack of D5-branes
which is wrapped on the central $\P^1$.
This is of particular importance for the geometric transition. After this
transition, the action (\ref{ofiveprime}) will generically become fixed point free
so the O5'-'plane' disappears, being replaced by the appropriate RR flux
on the three-cycle creating by smoothing the double point sitting at the origin.
This is quite different from the behavior of $O5$, which survives the transition as
we shall see in a moment.

\subsection{The T-dual configurations}

To extract the T-dual Hanany-Witten configurations, we again use a local
description valid on a subset $\tilde X_1 \subset \hat X_1$.  In the present case, it is given by two copies
$U_0$ and $U_1$ of $\C^3$ with coordinates $(x_i, u_i, z_i)$ ($i=0,1$) which  are glued together according to:
\be
(x_1, u_1, z_1) = (\frac{1}{u_0}, x_0 u_0^2 -2 W'(z_0) u_0, z_0)
\ee
The resolution map $\tau$ has the form:
\begin{eqnarray}
(z,u,x,y) =& ( z_0, x_0 u_0 - W'(z_0), x_0 , u_0(x_0 u_0 - 2 W'(z_0)) ) \,,\\
          =& ( z_1, x_1 u_1 + W'(z_1), x_1(x_1 u_1 + 2 W'(z_1)), u_1 )\,.
\end{eqnarray}
The $U(1)$ action (\ref{U1red}) is given by:
\be
\label{U1red_local}
(z_i, u_i, x_i ) \longrightarrow (z_i, e^{-i\theta}u_i, e^{i\theta}x_i)~~,
\ee
and fixes the rational curves $u_i=x_i=0$, which are the proper transforms of the two components
of (\ref{Sigma_10}).

The flat coordinates
of the Hanany-Witten construction are given by:
\be
x^4 + i x^5 = x_0 u_0 - W'(z_0)=x_1 u_1 +  W'(z_1)~~~,
~~x^6=\frac{1}{2}(|x_1|^2-|u_0|^2)~~
\label{base_coords_a1}
\ee
and $z=x^8+ix^9$, while $x^7$ is the periodic coordinate along the
orbits of the $U(1)$ action (\ref{U1red_local}). 

\begin{figure}[hbtp]
\begin{center}
 \scalebox{0.7}{\input{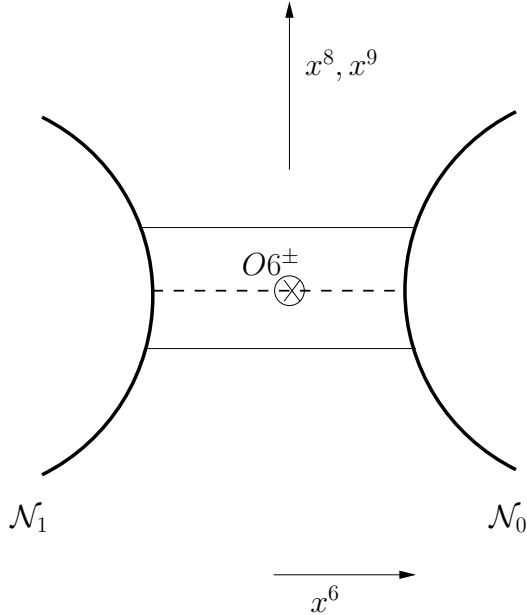}}
\end{center}
\caption{Brane construction with an O6 plane.}
\label{O6plane}
\end{figure}

As mentioned above, the T-dual background contains two NS5-branes ${\cal N}_0$ and ${\cal N}_1$,
which sit at
\bea
{\cal N}_0:~~x^4+ix^5=-W'(z)~~,~~x^6=+\infty
\eea
and:
\be
{\cal N}_1:~~x^4+ix^5=+W'(z)~~,~~x^6=-\infty~~.
\ee
The orientifold (\ref{red_or}) acts in local coordinates as:
\be
(z_0, x_0 , u_0) \longleftrightarrow (-z_1, -u_1, -x_1)\,.
\ee
This action fixes the locus $u_0^2+1=z=0$, which is a union of two
disjoint rational curves. 
Using (\ref{base_coords_a1}) we find that under T-duality this maps to
an O6-plane sitting at $x^6=x^8=x^9=0$ (figure \ref{O6plane}). Note
that there is a single dual O6-plane, even though the original
O5-plane in the resolved space ${\hat X}_1$ has two connected components. 
This is due to nonlinearity of the map (\ref{base_coords_a1}).

\begin{figure}[hbtp]
\begin{center}
  \scalebox{0.7}{\input{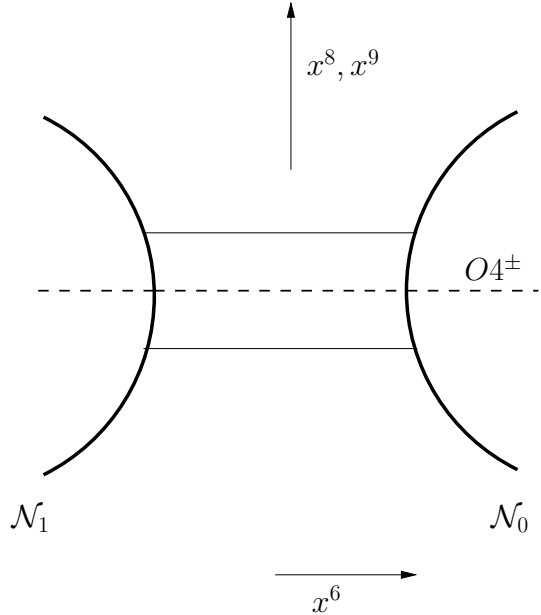}}
\end{center}
\caption{Brane construction with an O4 plane.}
\label{O4plane}
\end{figure}

On the other hand, the
orientifold (\ref{ofiveprime}) acts in local coordinates as:
\begin{eqnarray}
(z_0, x_0 , u_0) &\rightarrow& (-z_0, -x_0, u_0)~~~\mathrm{in~the~patch~U_0}\,,\\
(z_1, x_1 , u_1) &\rightarrow& (-z_1, x_1, -u_1)~~~\mathrm{in~the~patch~U_1}\,.
\end{eqnarray}
In the T-dual picture, this maps to an orientifold four-plane located at
$x^4=x^5=x^8=x^9=x^7=0$ (figure \ref{O4plane}).

For both choices of orientifold action, wrapping  D5-branes on the exceptional divisors will give
an ${\cal N}=1$ gauge theory with
orthogonal or symplectic gauge group depending on the RR-charge of the
orientifold 5-`plane'. The theory also contains a chiral multiplet $\Phi$ in the
adjoint representation of the gauge group with a tree-level superpotential
$tr W(\Phi)$. The classical vacua have the form:
\be
\label{phi_vev}
\Phi={\rm diag}(0_{N_0}, -\zeta_1 1_{N_1}, \zeta_1 1_{N_1} \dots -\zeta_n 1_{N_n}, \zeta_n 1_{N_n})
\ee
where $\zeta_j$ are the zeroes of $W'(z)$ and $N_0+2\sum_{j}{N_j}=N$. 
Here $j=-n\dots n$, with $\zeta_{-j}=-\zeta_j$ and in particular
$\zeta_0=0$. Such a vev of $\Phi$ breaks the gauge group down to the product:
\be
G_0 \times \prod_{j=1}^n{U(N_j)}~~,
\ee
where $G_0=SO(N_0)$ or $Sp(\frac{N_0}{2})$ according to whether we have an orientifold plane of positive or negative charge.

\subsection{Low energy descriptions after the geometric transition}

After the geometric transition, one must distinguish how one computes the effective superpotential in the
two constructions.

\subsubsection{Engineering through an O5-`plane' T-dual to an O4-plane}

Let us first review the results of
\cite{ashok} for the orientifold action (\ref{ofiveprime}). Since in this case the orientifold
is replaced by a contribution to the RR flux during the geometric transition, 
its effect is to modify the flux through the
three cycle ${\tilde A}_0$ of (\ref{a1_deformed}) which results by smoothing the central double point of $X_1$.
At low energies, the SO/Sp gauge
theory confines producing gaugino condensates ${\tilde S}_j$.
This corresponds to a geometric transition during which the exceptional
$\P^1$'s are blown down and the resulting singular space $X_1$ is smoothed to:
\be
\label{a1_deformed}
xy = u^2 - W'(z)^2 + 2 f_0(z)~~.
\ee
As usual, the deformation must be log-normalizable and respect the
orientifold projection, so $f_0(z)$ must be an even polynomial of degree $2n-2$.
We also included a factor of two in order to match the normalization that
arises naturally in the related matrix model.

The gaugino condensates ${\tilde S}_j$ can be identified with periods
of the holomorphic three-form of (\ref{a1_deformed}) over the
3-spheres produced by the transition. Using notation similar to that 
of Section 3,
the gaugino condensate ${\tilde S}_0$ in the $SO/Sp$ factor of the
unbroken gauge group
can be identified with the period of the holomorphic 
three-form $\Omega$ along the short invariant cycle $A_0$: 
\be
2 \pi i S_0 = \frac{1}{2}\int_{A_0} \Omega\,.
\ee
Then the RR-flux through the short invariant cycle $A_0$ is:
\be
\frac{1}{2}\int_{A_0} H = \frac{N_0}{2} -  s~~.
\ee
We find the flux superpotential:
\be
\label{ashok_eff}
W_{eff} = (\frac{N_0}{2} - s) 
\frac{\partial F_0}{\partial S_0}~~.
\ee
Note that there is no orientifold contribution because the
orientifold fixed `plane' is replaced by a RR flux after the
transition. In the corresponding matrix model there are of course still
diagrams with $\R\P^2$ topology and as shown in \cite{ino} they contribute
to the superpotential as:
\be
W_{eff} = \frac{N_0}{2}
\frac{\partial F_0}{\partial S_0} + 4 F_1~~,
\ee
where $F=F_0+\frac{1}{N} F_1 + O(1/N^2)$ is the microcanonical partition
function of the matrix model.  This allows one to identify the subleading 
$1/N$ contribution to the superpotential as \cite{ashok}:
\be
\label{their_description}
F_1 = -\frac{s}{4} \frac{\partial F_0}{\partial S_0}~~.
\ee

Reducing to the deformed Riemann surface:

\be
\label{a1_riemann}
u^2 = W'(z)^2 - 2 f_0(z)\,,
\ee
(the two-section $x=y=0$ of the deformed Calabi-Yau (\ref{a1_deformed})),
the relevant period integrals take the form:
\be
S_0 = \int_{a_0} u \frac{dz}{2\pi i} \,,~~~~\frac{\partial F_0}{\partial S_0} =
\Pi(\Lambda) = \int_{b_0(\Lambda)} u \,dz\,.
\ee
The `cycle' $b_0$ has intersection $-1$ with $a_0$ and is non-compact. 
The integral over $b_0$ has
been regularized by introducing a cutoff $\Lambda$.

\subsubsection{Engineering through an O5-`plane' T-dual to an O6-plane}

In this case, the transition again produces the space
(\ref{a1_deformed}), this time with the orientifold action
(\ref{red_or_proj}), which fixes the locus $z=0$, $x=-y$. This
is the smooth rational curve:
\be
x^2+u^2+2f_0(0)=0~~.
\ee 
Note that the O5-`plane' becomes connected after the geometric transition. 

To extract the effective superpotential, we shall use the trick of realizing
the theory with an O5-`plane' T-dual to an O6-plane as a certain Higgs branch of the theory of
with symmetric or antisymmetric matter. This branch is obtained from (\ref{phi_vev_sa}) when taking
$W(z)$ to be even of degree $d+1=2n$, which forces the $d-1=2n-2$ critical points $z_j$ of $W$
to coincide with the $2\delta=2n-2$ nonvanishing solutions ${\tilde z}_j$ of $W'(z)-W'(-z)=0$.
With an appropriate enumeration, we
can then take ${\tilde z}_j=z_j^+$ for all positive $j$ and ${\tilde z}_{j}=z_{-j}^-$ for all negative $j$.
We also have the null value ${\tilde z}_0=0$. Then one can further Higgs by giving nonzero expectation
values to $Q$, which forces us to keep only solutions of type (\ref{phi_vev}), thus recovering
the vacua of the SO/Sp theory.

In the brane construction, this process amounts to displacing the central
NS-brane in order to give vevs to Q, which eliminates all stacks of D4 branes
stretching between the middle and outer
NS branes. Then the 
middle NS brane can be decoupled, which recovers the realization which
uses only the two outer NS branes. 

In geometric engineering, this corresponds to starting with the special case when
the classical curve (\ref{Sigma_0}) is reduced to the form:
\be
\label{Sigma_0_special}
\Sigma_0^{special}:~u(u-W'(z))(u+W'(z))=0~~,
\ee
which has {\em triple} points at the critical points of $W$. Then giving a vev to $Q$ amounts to forgetting the
branch $u_2\equiv 0$, thereby recovering the classical SO/Sp curve (\ref{Sigma_10}). Starting with such special
orientifolded $A_2$ fibrations, the geometric transition will produce a deformed space $X$ which is only a
{\em partial} smoothing of $X_0$. Namely, the $A_2$ singularities of $X_0$ are deformed into $A_1$ singularities,
which corresponds to partially smoothing the triple points of (\ref{Sigma_0_special}) by replacing the 
factor (\ref{Sigma_10}) with its deformation (\ref{a1_riemann}).

In the matrix model, this amounts to requiring that all filling fractions $S_j$ must vanish,
so that all cuts connecting the branches
$u_0$ and $u_2$ as well as $u_0$ and $u_2$ are reduced to double
points. As we shall see in a moment, this amounts to restricting to planar eigenvalue
distributions which are symmetric with respect to the origin of the $z$-plane. 

To see this explicitly, remember from \cite{us} that the large $N$ spectral curve of the theory 
with symmetric or antisymmetric matter (and without the
logarithmic deformation) has the form (\ref{Sigma}), where
the polynomials $p,q$ have the following expression in terms of matrix model data:
\bea
\label{pq}
p(z)&=& t(z)^2+t(z)t(-z)+t(-z)^2-f_0(z)-f_0(-z)~~\nn\\
q(z)&=& -t(z)t(-z)\left[t(z)+t(-z)\right]+t(z)f_0(-z)+t(-z)f_0(z)-g_0(z)-g_0(-z)~~.
\eea
Here $f_0(z)$ and $g_0(z)$ are polynomials of degree $d-1= 2n-2$, given explicitly by:
\bea
\label{fg}
f_0(z)&=&\int{d\lambda \rho_0(\lambda)\frac{U'(z)-U'(\lambda)}{z-\lambda}}\nn\\
g_0(z)&=&\int{d\mu \int{d\lambda \rho_0(\lambda) \frac{U'(z)-U'(\lambda)}{(\lambda+\mu)(z-\lambda)}}}~~
\eea
where $U'(z)=W'(z)+\frac{s}{2N}\frac{1}{z}$.

In the case of interest for this section, we have $W'(-z)=-W'(z)$ so that $U'(-z)=-U'(z)$ and $t(-z)=-t(z)=W'(z)$.
Since $deg W'(z)=d=2n-1$, we have $\delta=\left[\frac{d-1}{2}\right]=n-1$ and the surface (\ref{Sigma}) has
$2d=2(n-1)$ cuts of type $I_j$ and $2\delta+1=2n-1$ cuts of type ${\tilde I}_j$.

Let us assume that $\rho_0(-\lambda)=\rho_0(\lambda)$ for all $\lambda$. Then equations (\ref{fg}) immediately
imply that $f_0(-z)=f_0(z)$ and $g_0(-z)=-g_0(z)$, so that (\ref{pq}) give:
\bea
\label{pq_special}
p(z)&=& W'(z)^2-2f_0(z)\nn\\
q(z)&=& 0~~.
\eea
Therefore, the spectral curve (\ref{Sigma}) reduces to:
\be
\label{Sigma_special}
\Sigma_{special}:~~u^3-(W'(z)^2-2f_0(z))u=0\Longleftrightarrow u(u-W'(z)^2-2f_0(z))=0~~{\rm with}~~f_0={\rm even}~~.
\ee

Conversely, let us assume that (\ref{Sigma}) has the form (\ref{Sigma_special}). Then $u_2\equiv 0$ and
$u_1(z)=u_0(-z)=-u_0(z)$, which means that all cuts $I_j$ are reduced to ordinary double points
sitting at the endpoints of the cuts ${\tilde I}_j$ (there are two such double points for each cut ${\tilde I}_j$).
These double points correspond to the zeroes of the
degree $2(2n-1)$ even polynomial $W'(z)^2-2f_0(z)$, which are distributed symmetrically with respect to the origin
of the $z$-plane.

Now remember from Section 3.5.2 of \cite{us} (see equation (3.81) of that paper)
that the planar spectral density $\rho_0(\lambda)$ is symmetric along the union of all cuts of type
${\tilde I}_j$. Since in our case there are no other cuts, the support of $\rho_0$ coincides with $\cup_j{\tilde I}_j$
and it immediately follows that $\rho_0(-\lambda)=\rho_0(\lambda)$ for all $\lambda$. Thus:

{\em The large $N$ spectral density $\rho_0(\lambda)$
of the matrix model for symmetric or antisymmetric matter with an even tree-level superpotential $W$ of degree $2n$
is symmetric if and only if
the spectral curve has the form (\ref{Sigma_special}), i.e. iff. the polynomials $p(z)$ and $q(z)$ have the form
(\ref{pq_special}), where $f_0(z)$ is an {\em even} polynomial of degree $2n-2$.}

Hence the field theory higgsing described above translates into imposing a symmetric distribution for
$\rho_0(\lambda)=\lim_{N\rightarrow \infty}{\langle
  \rho(\lambda)\rangle}$. In this case the spectral curve
reduces to the form (\ref{Sigma_special}), which coincides (up to the
spectator branch $u\equiv 0$) with the spectral curve
(\ref{a1_riemann}) of the $SO/Sp$ model. It follows that the planar free
energy on this branch of the moduli space of filling fractions of the model with symmetric/antisymmetric
matter must agree with that of the SO/Sp model, provided that one
identifies the filling fractions ${\tilde S}_j$ of the former with those
of the latter. One the other hand, the
$\R\P^2$ contributions are given by relations (\ref{F1S0}) and (\ref{their_description}),
which have the same form and determine $F_1$ in terms of $F_0$. It follows
that the subleading terms $F_1$ must also agree.

\paragraph{Observation}

If one formally restricts to exactly symmetric distributions of eigenvalues of
the model with symmetric or antisymmetric matter, then   
one has the `quantum' relation $\omega(z)=-\omega(-z)$. Combining this
with $U'(z)=-U'(-z)$, the
quadratic loop equation in (\ref{ex_loop}) reduces to:
\be
\label{loop.eqn.sosp}
\frac{1}{2} \langle \omega(z)^2\rangle = \int_{\gamma} \frac{dx}{2\pi
  i} \frac{U'(x)}{z-x} \langle \omega(x) \rangle~~
\ee
while the cubic loop equation becomes a tautology. To arrive at (\ref{loop.eqn.sosp}), we
  used the identity:
\be
\frac{1}{z}\left(\frac{1}{z-\lambda_i}+\frac{1}{z+\lambda_i}\right) =
\frac{1}{\lambda_i}\left(\frac{1}{z-\lambda_i}-\frac{1}{z+\lambda_i}\right)=
\frac{2}{z^2-\lambda^2}~~.
\ee
Since the eigenvalues are distributed symmetrically, the resolvent takes
the form $\omega(z) = \sum_i \frac{2z}{z^2 - \lambda_i^2}$
\cite{ashok}.
Using $U'(x) = W'(x) + \frac{s}{2N}\frac{1}{x}$, we can write (\ref{loop.eqn.sosp}) as:
\be
\frac{1}{2}\langle \omega(z)^2 \rangle - \frac{s}{2N} \frac{1}{z}
\langle \omega(z) \rangle = \int_{\gamma} \frac{dx}{2\pi i}
\frac{W'(x)}{z-x} \langle \omega(x) \rangle\,.
\ee
This coincides with the exact loop equations for the SO/Sp theory with
adjoint matter, as derived in \cite{janik, ashok}.

Notice however that such a symmetric distribution
of eigenvalues might be unstable to orders $O(1/N^2)$, and a priori
there is no meaningful way to impose the symmetry of eigenvalues at
the quantum level in the matrix model (as opposed to symmetry of their
(large $N$) averaged distribution $\langle \rho(\lambda)\rangle$, which is all we
required before). Indeed, one expects that gravitational F-terms in
the effective field theory should distinguish between the two
realizations. In the matrix model, such 
gravitational corrections will correspond \cite{DV} to contributions
of order $O(1/N^2)$ . Therefore we expect that the two backgrounds
obtained after the large $N$ transition differ through F-terms of order
$O(1/N^2)$ or higher.

\section{Conclusions}
\label{conclusions}

We investigated the geometric engineering of ${\cal N}=1$ gauge theory
 with gauge group $U(N)$ and matter in the adjoint and symmetric or
 antisymmetric representations. We showed that such theories 
can be realized as certain orientifolds of resolved Calabi-Yau $A_2$
 fibrations, where one wraps $D5$-branes on the exceptional $\P^1$'s
 of the resolution. The orientifold action one has to consider defines
 an orientifold 5-plane whose internal part coincides with a
 {\em noncompact} rational curve sitting in the resolved Calabi-Yau
 space. 

We also gave the explicit relation of this
 construction with the Hanany-Witten realization  \cite{Karl_nonchiral0, Karl_nonchiral1}
 through orientifolded brane configurations in IIA string theory.
This is implemented  by T-duality with respect to the orbits of a
 certain $U(1)$ action on the resolved Calabi-Yau, along the lines of
\cite{OT,OT1,OT2}. Upon giving an explicit construction of the T-dual
 coordinates, we showed that the orientifold 5-`plane'
 used in our geometric engineering maps to the O6-plane used in
 \cite{Karl_nonchiral0, Karl_nonchiral1}.

Following the ideas of \cite{civ, Cachazo_Vafa, Cachazo_Vafa_more},
we considered the geometric transition which replaces the resolved
Calabi-Yau space with its deformation. Upon restricting to
deformations compatible with the geometric symmetry used in the
orientifold construction, we found that the orientifold 5-`plane' {\em
  survives} the geometric transition, and thus contributes to the
effective superpotential of the resulting background. This gives the
geometric explanation of the $\R\P^2$ diagram contributions found in
\cite{us}, which affect the field theory glueball superpotential after confinement. 

Using the matrix model results of \cite{us}, we gave a geometric
expression for this orientifold contribution as an integral over a
certain 3-chain in the deformed Calabi-Yau space, and compared with
the proposal of \cite{Vafa_or}. 

We also discussed the Higgs branch of our theories obtained by giving a vev to the symmetric or
antisymmetric tensor, and showed that this Higgs branch recovers the $SO(N)$ or $Sp(N/2)$ gauge theory
with adjoint matter. Namely, we showed that this process leads to a
geometric engineering of such theories which is T-dual to their
realization obtained by adding an O6-plane to a Hanany-Witten brane
configuration. 
Using the matrix model results of \cite{us}, we extracted the
glueball superpotential on this branch, and showed that it recovers
the results of \cite{ashok}. 
The later where obtained in \cite{ashok} though a different geometric engineering
(which is T-dual to a Hanany-Witten construction involving an
O4-plane). After the geometric transition, this alternate construction
leads to a pure flux background, in which the O5-`plane' is replaced
by R-R fluxes. This implies matching of $\R\P^2$ contributions to the
superpotential between the two constructions, which we checked explicitly
by using the matrix model results of \cite{us} and \cite{ashok}.

One can apply similar methods to other models
which admit a geometric engineering. 
An interesting example of this kind is the chiral $U(N)$ model with
adjoint as well as symmetric, antisymmetric and
fundamental matter which is discussed in \cite{chiral}.

\acknowledgments{ This work was supported by DFG grant KL1070/2-1. R.
  T. was supported in part by the DOE Contract DE-AC03-76SF00098 and
  in part by the NSF grant PHY-0098840.}

\appendix

\section{Geometric engineering without a tree-level superpotential}
\label{without}

In this appendix we give a geometric argument which explains why the orientifold
projection of our IIB background leads to (anti)symmetric matter.
For this, one can consider the limit when the tree-level superpotential
vanishes and the theory acquires ${\cal N} = 2$ supersymmetry. In this limit,
the Calabi-Yau space $X_0$ becomes the trivial $A_2$ fibration $\C\times X_0(0)$,
where the fiber $X_0(0)$ is an $A_2$ singularity.

More generally, the relevant geometry can be obtained by considering the non-generic case when
$W'(0)=0$ (note that this case was explicitly excluded in Section 2).
Then the central fiber $X_0(0)$ becomes an $A_2$ singularity
$xy=u^3$, while ${\hat X}(0)$ becomes its minimal resolution: \bea
\label{a2}
\xi_1x&=&u\nn\\
\xi_2u&=&y\nn\\
\frac{\xi_2}{\xi_1}&=&u\\
xy&=&u^3\nn~~.  \eea Outside the locus $x=y=u=0$, (\ref{a2})
determines\footnote{When $xy=0$, but $x\neq y$, one uses one of the
  forms given in the text to remove the ambiguity. For example, let
  $x=0$ and $y\neq 0$. Then one must use the forms
  $\xi_1=\frac{y}{u^2}$ and $\xi_2=\frac{y}{u}$ in order to determine
  $\xi_1$ unambiguously as a point in the one-point compactification
  ${\overline \C}=\P^1$. } $\xi_1=\frac{u}{x}=\frac{y}{u^2}$ and
$\xi_2=\frac{y}{u}=\frac{u^2}{x}$. For $x=y=u=0$, we are left with the
constraint $\frac{\xi_2}{\xi_1}=0$, which gives the two exceptional
$\P^1$'s with equations $\xi_1=\infty$ and $\xi_2=0$. The orientifold
action on ${\hat X}$ fixes the locus ${\hat O}: x=-y, \xi_1\xi_2=-1$
in ${\hat X}(0)$, which is a a smooth rational curve passing through
the common point $p$ of the two exceptional $\P^1$'s: \be
\label{point}
D_0^{(1)}\cap D_0^{(2)}=\{p\}: x=y=0,~\xi_1=\infty,~\xi_2=0~~.  \ee
The curve ${\hat O}$ intersects the exceptional fibers only at this
point, and the orientifold action maps $D_0^{(1)}$ into $D_0^{(2)}$
according to relation (\ref{xi_action}). Finally, note that ${\hat
  O}$ projects to the fixed locus $O_0$ of the action
(\ref{or_projected}) on $X_0$, which in the case $W'(0)=0$ is the
cuspidal curve: \be
\label{cusp}
x=-y,~~x^2+u^3=0~~.  \ee

A clearer description of the resolved central fiber ${\hat X}(0)$ presents it as the toric
resolution $(\C^4-Z)/(\C^*)^2$ of the $A_2$ singularity $X_0(0)$, with
charge matrix: \be
\label{charge}
Q=\left[\begin{array}{cccc}1&-2&1&0\\0&1&-2&1\end{array}\right]~~ \ee
and toric generators given by the columns of the matrix: \be
G=\left[\begin{array}{cccc}1&1&1&1\\0&1&2&3\end{array}\right]~~.  \ee
The exceptional set is $Z=\{x_1=x_3=0\}\cup \{x_2=x_4=0\}\cup
\{x_1=x_4=0\}$.

The generators correspond to homogeneous coordinates which we denote
by $x_1\dots x_4$.  We let $D_j=(x_j)$ be the toric divisors.  Then
$D_2= D_0^{(1)}$ and $D_3=D_0^{(2)}$ are the exceptional $\P^1$'s,
while $D_1$ and $D_4$ are non-compact.

\begin{figure}[hbtp]
\begin{center}
  \scalebox{0.7}{\input{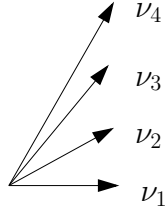}}
\end{center}
\caption{The toric generators $\nu_1\dots \nu_4$.}
\label{gens}
\end{figure}

In the symplectic quotient description, we have the reduction of
$\C^4$ with respect to the $U(1)^2$ action defined by (\ref{charge})
with moment map equations: \bea
|x_1|^2-2|x_2|^2+|x_3|^2&=&\zeta_1\nn\\
|x_2|^2-2|x_3|^2+|x_4|^2&=&\zeta_2~~, \eea where $\zeta_j$ are some
positive levels. Setting $\zeta_1=\zeta_2=0$ gives the $A_2$
singularity $X_0(0)$, which we describe in terms of invariants: \bea
x&=&x_1^3 x_2^2x_3\nn\\
y&=&x_2x_3^2x_4^3\\
u&=&x_1x_2x_3x_4~~,\nn \eea subject to the relation $xy=u^3$.

\begin{figure}[hbtp]
\begin{center}
  \scalebox{0.7}{\input{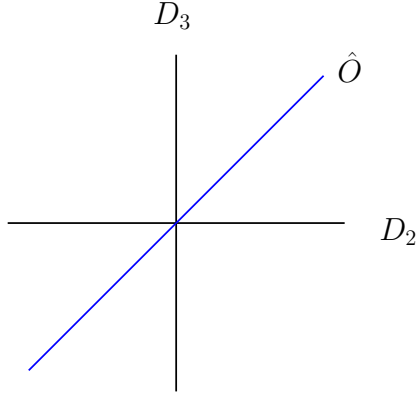}}
\end{center}
\caption{Orientifold action on the resolution of the $A_2$ singularity.}
\label{fiber_or}
\end{figure}

The orientifold action on ${\hat X}(0)$ takes the form \footnote{In
  the symplectic quotient description, this is a symmetry if we set
  $\zeta_1=\zeta_2=\zeta$.} : \be
\label{Yaction}
x_1\longleftrightarrow x_4~~,~~x_2\longleftrightarrow -x_3~~.  \ee The fixed point set
${\hat O}$ is given by the smooth curve $x_1^3x_2=-x_3x_4^3$.
One easily checks the intersections: \be D_2\cap D_3=D_2\cap {\hat
  O}=D_3\cap {\hat O}= D_2\cap D_3\cap {\hat O}=\{[1,0,0,1]\}~~.  \ee
The curve ${\hat O}$ is smooth for positive $\zeta$ and degenerates to
the cusp $O_0$ of (\ref{cusp}) when $\zeta=0$.  As expected, the
action (\ref{Yaction}) permutes the compact divisors $D_2=D_0^{(1)}$
and $D_3=D_0^{(2)}$. The local geometry is sketched in figure
\ref{fiber_or}.

It is now trivial to find the orientifolded matter content. In the
limit $W'(z)=0$, we start by wrapping two D5-branes on the exceptional
$\P^1$'s $D_0^{(1)}$ and $D_0^{(2)}$, endowed with trivial and
isomorphic Chan-Paton bundles $E_1$ and $E_2$ of rank $N$.
This generates chiral multiplets $Q_{21}\in
Hom(E_1|_p, E_2|_p)\approx Mat(N,\C)$ and $Q_{12}\in Hom(E_2|_p,
E_1|_p)\approx Mat(N,\C)$ from the intersection point $p$, as well as
gauge multiplets from the strings ending on a given brane.
In ${\cal N}=1$ superfield language the strings ending on a
brane give rise to vector fields  and chiral
multiplets in the adjoint representation. More precisely we have
$\W_{\alpha.j} \in Aut(E_{j,p}) \approx Mat(N,\C)$ and
$\Phi_j \in Aut(E_{j,p}) \approx Mat(N,\C)$ for $j\in \{1,2\}$, where
$\Phi_j$ correspond to moving the brane along the base direction $z$ in
the trivial fibration $\C\times X(0)$.

The result
is an $A_2$ quiver field theory whose node potentials $W_1$ and $W_2$
arise when one deforms the trivial fibration $\C\times {\hat X}(0)$ in
order to obtain the nontrivial fibration ${\hat X}$ over the $z$-plane.

Let us now add the orientifold 'plane'.  Since the orientifold action
permutes the two $\P^1$'s, the projection must relate each of $Q_{12}$ and
$Q_{21}$ to its transpose, up to a similarity defined by
homomorphisms $\gamma_1\in Hom(E_1|_p, E_2|_p)$ and $\gamma_2\in
Hom(E_2|_p, E_1|_p)$: \bea
 Q_{ij} \rightarrow \gamma_i Q_{ij}^T \gamma_j^{-1}
\eea
Since this action must square to the identity, we have the
constraints: \be \gamma_2=\pm \gamma_1^T~~.  \ee

To recover our theories, we choose $\gamma_1=+1_N$ and $\gamma_2=\pm
1_N$, which give respectively the projections:
\be Q_{ij}^T=+Q_{ij}~~\mathrm{or}~~
  Q_{ij}^T=-Q_{ij}~~.  \ee
These are precisely the projections used in the
introduction, corresponding to the two choices $s=\pm 1$.
Finally notice that on the vector multiplets and the chiral
multiplets in the adjoint representation the orientifold projection
acts as
\bea
\W_{\alpha , 1}  & \rightarrow &
- \gamma_1 \W_{\alpha ,2} ^T \gamma_1^{-1}  \,,\\
\W_{\alpha , 2}  & \rightarrow &
- \gamma_2 \W_{\alpha,1}^T \gamma_2^{-1}  \,,\\
\Phi_1 & \rightarrow &
- \gamma_1 \Phi_2^T \gamma_1^{-1}  \,,\\
\Phi_2 & \rightarrow &
- \gamma_2 \Phi_1^T \gamma_2^{-1}  \,.
\eea
The additional sign arises for the vector fields because the
vertex operators of vectors are odd under worldsheet parity, whereas
the minus sign for the chiral multiplets has its origin in the
geometric action of the orientifold. According to these projections
the two factor groups of the $A_2$ quiver gauge theory are identified
as $U_1 = (U_2^T)^{-1} = U_2^*$ for both choices of $\gamma_i$.
Writing now $\Phi=\Phi_1=-\Phi_2^T$, $Q=Q_{21}$ and $\tilde Q = Q_{12}$
we can compute the projected superpotential (\ref{quiver_superpot})
\be
W = \tr(Q \Phi \tilde Q) - \tr ( \tilde Q (-\Phi^T) Q) =
 2\, \tr(Q \Phi \tilde Q )\,.
\ee
The factor $2$ can be absorbed in the normalization of $Q$ and $\tilde Q$
to produce (\ref{symasym_superpot}).

\end{document}